\documentclass[sigconf,screen]{acmart} 

\usepackage{natbib}
\usepackage{xcolor}  
\usepackage[normalem]{ulem}  
\usepackage{soul}       
\usepackage[dvipsnames]{xcolor}
\usepackage{algorithm}
\usepackage{algorithmic}
\usepackage{balance}
\usepackage{placeins}
\usepackage{afterpage}

\usepackage{listings}
\usepackage[utf8]{inputenc}
\usepackage{tcolorbox}
\usepackage{threeparttable}
\usepackage{pifont}
\usepackage{subcaption}
\usepackage{xurl}
\makeatletter
\g@addto@macro\UrlBreaks{\do\_\do\-\do\/}
\makeatother

\newcommand{\codeid}[1]{%
  \begingroup
  \def\UrlFont{\ttfamily}%
  \nolinkurl{#1}%
  \endgroup
}

\AtBeginDocument{%
  }

\setcopyright{acmlicensed}
\copyrightyear{2018}
\acmYear{2018}
\acmDOI{XXXXXXX.XXXXXXX}

\acmConference[Conference acronym 'XX]{Make sure to enter the correct
  conference title from your rights confirmation email}{June 03--05,
  2018}{Woodstock, NY}

\acmISBN{978-1-4503-XXXX-X/2018/06}

\newcommand{\Cg}{\textit{CogInstrument}}

\renewcommand\hl[1]{#1} 

\begin{document}

\title[\textit{CogInstrument}]{CogInstrument: Modeling Cognitive Processes for Bidirectional Human-LLM Alignment in Planning Tasks}



\author{Anqi Wang}
\orcid{0000-0003-4238-6581}
\affiliation{%
  \institution{Hong Kong University of Science and Technology}
  \city{Hong Kong SAR}
  \state{}
  \country{China}
}

\author{Dongyijie Pan}
\orcid{0009-0005-9830-1614}
\affiliation{%
  \institution{Hong Kong University of Science and Technology (Guangzhou)}
  \city{Guangzhou}
  \country{China}}

\author{Xin Tong}
\orcid{0000-0002-8037-6301}
\affiliation{%
  \institution{Hong Kong University of Science and Technology (Guangzhou)}
  \city{Guangzhou}
  \country{China}}

\author{Pan Hui}
\authornote{Corresponding author}
\orcid{0000-0001-6026-1083}
\affiliation{%
  \institution{Hong Kong University of Science and Technology (Guangzhou)}
  \city{Guangzhou}
  \country{China}
  \email{panhui@hkust-gz.edu.cn}
}
\affiliation{
  \institution{Hong Kong University of Science and Technology}
  \city{Hong Kong SAR}
  \country{China}
  \email{panhui@ust.hk}
}

\begin{abstract}



Although Large Language Models (LLMs) demonstrate proficiency in knowledge-intensive tasks, current interfaces frequently precipitate cognitive misalignment by failing to externalize users' underlying reasoning structures. Existing tools typically represent intent as "flat lists," thereby disregarding the causal dependencies and revisable assumptions inherent in human decision-making. We introduce \Cg{}, a system that represents user reasoning through cognitive motifs—compositional, revisable units comprising concepts linked by causal dependencies
. 
\Cg{} extracts these motifs from natural language interactions and renders them as editable graphical structures to facilitate bidirectional alignment. This structural externalization enables both the user and the LLM to inspect, negotiate, and reconcile reasoning processes iteratively. 
A within-subjects study ($N=12$) demonstrates that \Cg{} explicitly surfaces implicit reasoning structures, facilitating more targeted revision and reusability over conventional LLM-based dialogue interfaces. 
By enabling users to verify the logical grounding of LLM outputs, \Cg{} significantly enhances user agency, trust, and structural control over the collaboration. This work formalizes cognitive motifs as a fundamental unit for human–LLM alignment, providing a novel framework for achieving structured, reasoning-based human–AI collaboration. 


\end{abstract}

 

\begin{teaserfigure}
\centering
   \includegraphics[width=\textwidth]{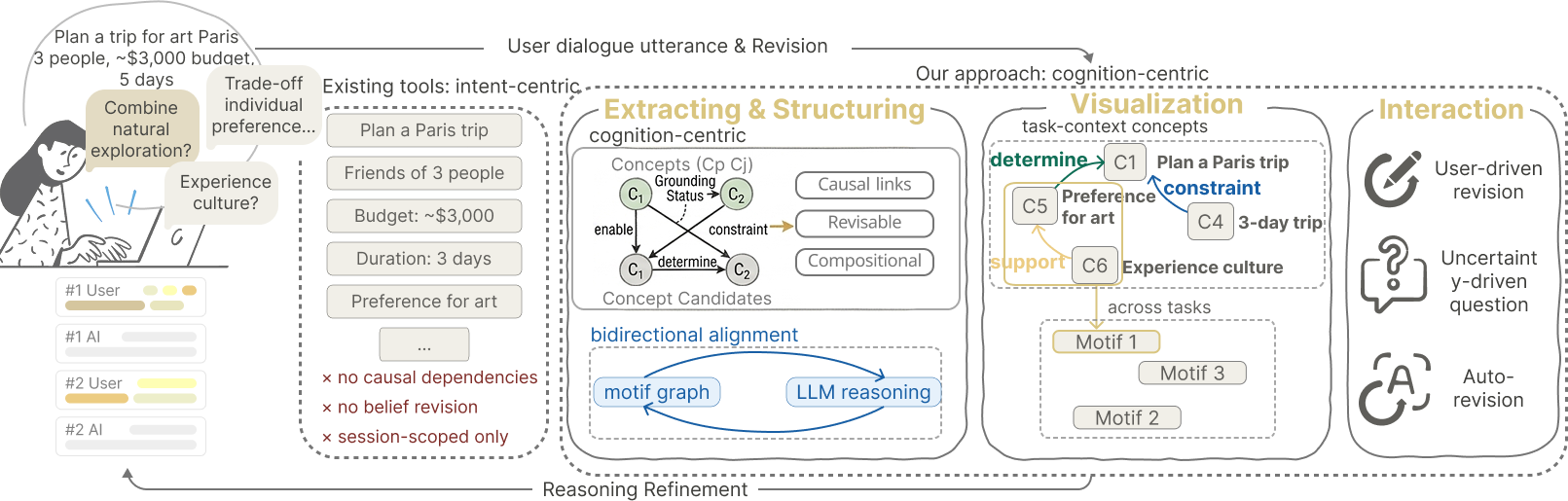}
   \caption{
   From intent-centric prompting to cognition-centric interaction. We introduce a pipeline that extracts and structures users’ cognitive processes (e.g., belief, preference, constraints, and dependencies) into manipulable \emph{cognitive concept} and \emph{motif} graphs, enabling visualization and iterative interaction. This reification supports bidirectional alignment between human reasoning and LLM behavior.}
   \Description{Pipeline of CogInstrument}
  \label{fig:teaser}   
\end{teaserfigure}

\received{20 February 2007}
\received[revised]{12 March 2009}
\received[accepted]{5 June 2009}


\maketitle

\section{Introduction}
Recent advances in large language models (LLMs) have transformed user engagement in knowledge-intensive tasks, including planning \cite{kim2025plantogether,suh2023sensecape}, design \cite{wang_2026_designerlyloop,suh_luminate_2024,choi_creativeconnect_2024}, writing \cite{guo_pen_2025,zhang_synthia_2025}, and programming \cite{zhou_instructpipe_2025,feng_coprompt_2024}. This paper focuses on planning as a representative domain in which users must articulate goals, negotiate constraints, revise assumptions, and reuse reasoning over time. As users interact with LLMs, they continuously form intentions, generate hypotheses, anticipate outputs, and evaluate results \cite{direct1985hutchins,hutchinsCognitionWild1995,norman_cognitive_1986}. Yet, current human--LLM interfaces provide limited support for making these cognitive processes explicit, inspectable, and structurally revisable during interaction \cite{subramonyam_bridging_2024}. Much of this reasoning remains tacit: users cannot readily inspect or organize evolving goals, while LLMs lack access to the underlying logic of user prompts \cite{li2025simulating,subramonyam_bridging_2024}. Consequently, human--LLM interaction often devolves into trial-and-error, increasing the risk of reasoning misalignment and outputs that diverge from user intentions \cite{10.1145/3706598.3713166,10.1145/3613904.3642466}. To address this, cognitive alignment is essential—align interaction around user-grounded reasoning structures beyond merely to match final outputs \cite{amershi2014power}. This approach grounds LLM outputs in actual reasoning rather than surface-level text \cite{li2025simulating}, enabling users to revise that reasoning during interaction \cite{mcneil_visualizing_2015} and fostering bidirectional trust \cite{yin_operation_2025,zhang_neurosync_2025}. 
Existing approaches address this gap only partially and fail to represent it structurally. Intent-externalization tools—such as IntentFlow~\cite{kim_intentflow_2025}, OnGoal~\cite{coscia_ongoal_2025}, and Intent Tagging~\cite{gmeiner_intent_2025}—capture user intent as flat lists or shallow hierarchies, representing the outcomes of cognition rather than its generative structure. Human decision-making, however, is rarely reducible to a flat list; it is organized through causal dependencies among assumptions, conditional schemas, and beliefs whose implications shift upon revision. Flat representations cannot encode these dependencies, detect conflicts, or support principled revision when underlying beliefs change.
~
By contrast, model-side reasoning externalization tools—such as NeuroSync~\cite{zhang_neurosync_2025}—visualize the model's internal reasoning as an editable graph. While complementary, this approach addresses a different problem: it centers on the model's interpretation rather than the user's reasoning, assuming users possess fully formed intentions. We formalize this gap as cognitive misalignment: a mismatch between the structured, causally organized nature of user reasoning and the flat, session-scoped representations of current interfaces. 

Thus, our work explores a cognition-centric approach to human--AI alignment that explicitly operates on user cognitive structures by externalizing reasoning as shared, inspectable approach for coordination. 
Drawing on cognitive science~\cite{gopnik2004causal,pearl2009} and LLM reasoning literature~\cite{li2025simulating}, we propose cognitive \textbf{motifs}: a small taxonomy of recurring reasoning patterns that capture at least cognitive \textbf{concepts} of human reasoning and their causal links. A motif is a compositional, revisable unit—for example, $group\_size \to accommodation\_type \to budget\_allocation$. Grounded in this formulation, we implement \textbf{\Cg{}} (Figure~\ref{fig:teaser}), a system that: (1) extracts cognitive concepts and motifs from user interactions as reusable structures; (2) externalizes these as editable graphs for direct inspection and revision; and (3) maintains a shared reasoning space for iterative negotiation between the user and the LLM. Alignment is bidirectional: the LLM is grounded in the reconstructed graph, and its outputs are checked against active motifs to reveal which elements of user reasoning were incorporated, ignored, or mismatched. 
Our aim is to externalize and revise user reasoning's causal links, rather than model full human cognition. 

A within-subjects study ($N=12$) demonstrates that \Cg{}’s motif-based representation facilitates superior articulation and diagnosis of complex reasoning over conventional LLM-based dialogue. This causal transparency enables precise editing—reducing repair effort—while fostering higher trust and user agency by making the model's logic verifiable beyond surface-level text. 
~
Our work offers three contributions: (1) A theoretical formalization of cognitive motifs as fundamental units for human–LLM alignment, grounded in the causal, compositional, and revisable nature of human reasoning; (2)  The \Cg{} system, an interaction approach that extracts motifs from natural language and renders them as manipulable graphs to facilitate bidirectional human–AI revision; and (3) An empirical study ($N=12$) demonstrating that motif-based externalization facilitates the diagnosis of complex reasoning and provides a novel analytical lens for studying human–AI alignment. 

\section{Related Works}

\subsection{Externalizing Cognitive Processes in HCI}\label{sec:rw-cognition}
Prior work in cognitive and learning sciences suggests that human reasoning is structured rather than flat: it involves causal dependencies, compositional reuse, and ongoing revision \cite{norman_cognitive_1986,norman_design_2013,gopnik2004causal,julian_naive_2020,goodman2014concepts,wong2023wordm,lake_2017_building,tenenbaum_2011_growmind,willemain_visualization_2019}. HCI has long explored how to externalize such latent structures to bridge users’ mental intentions and system actions through graphical or linguistic artifacts \cite{olson1995growth,duric2002integrating,salvucci_simple_2003,ifenthaler_identifying_2011,tauber_mental_1991,argyris_teaching_2002,subramonyam_bridging_2024,wang_mental_2025,son_clearfairy_2025,yin_operation_2025,tankelevitch_metacognitive_2024,zhang_synthia_2025}. Systems such as TaskMind and ClearFairy formalize decision structures or task graphs from user behavior, while CausalMapper uses causal maps and LLM prompting to support mixed-initiative reflection \cite{yin_operation_2025,son_clearfairy_2025,huang_causalmapper_2023}. These systems show that externalization can improve reflection and coordination, but they do not provide a shared representation for inspecting and revising user-grounded reasoning across tasks \cite{tankelevitch_metacognitive_2024}.

\subsection{Human--LLM Alignment}

Recent work on human--LLM alignment addresses the gap between internal mental states and model outputs~\cite{zhang_neurosync_2025,shaikh_creating_2025,ma_deliberation_2025}. Most existing systems, however, externalize either user-side goals or model-side interpretations, while leaving underspecified the reasoning structure that connects them.

\emph{Intent Externalization: Capturing ``What'' Users Want} 
A growing body of work makes user intent more explicit and manipulable. Object- and component-oriented frameworks such as \textit{Cells, Generators, and Lenses}~\cite{kim_cells_2023} and \textit{Brickify}~\cite{shi_brickify_2025} decompose prompts into reusable parts that users can steer directly. 
\textit{OnGoal}~\cite{coscia_ongoal_2025}, \textit{Intent Tagging}~\cite{gmeiner_intent_2025}, \textit{IdeaBlocks}~\cite{choi_ideablocks_2025}, and \textit{AI-Instruments}~\cite{riche_ai-instruments_2025} further turn intents into manipulable annotations or visual objects. Other work reduces explicit specification burdens by inferring user knowledge or preferences from behavior, as in \textit{InterQuest}~\cite{interQuest_2025_mei} and the General User Model (GUM)~\cite{shaikh_creating_2025}. Recent work such as \textit{Semantic Commit}~\cite{vaithilingam_semanticcommit_2025} further highlights that once intent specifications persist over time, users also need support for updating and reconciling them without losing semantic continuity.


\emph{Model-Side Reasoning Externalization: Visualizing ``How'' Models Interpret Tasks} 
A complementary line of work helps users anticipate or inspect model behavior. \textit{NeuroSync}~\cite{zhang_neurosync_2025} responds by externalizing the LLM's task interpretation as an editable graph before code generation. \textit{TaskMind}~\cite{yin_operation_2025} and \textit{DesignerlyLoop}~\cite{wang_2026_designerlyloop} similarly surface model-side structures to support task-specific steering and correction. Related work on human--AI deliberation likewise seeks to make model-supported decision processes more discussable and revisable during interaction~\cite{ma_deliberation_2025}.

\emph{Research Gap and Our Position}
Subramonyam et al. \cite{subramonyam_bridging_2024} identify a ``gulf of envisioning,'' wherein users must simultaneously articulate intent and anticipate the behavior of probabilistic models. 
Despite current works attempts to bridge this gap, both lines generally failed: Intent-externalization systems capture \emph{what} users want more effectively; model-side reasoning systems only improve visibility into how the model parses a task. Both lines of these systems omit giving users a shared structure for their own reasoning, such as assumptions, constraints, evidence, and trade-offs behind decision or intents. 

Thus, this work aims to propose a paradigm shift from intent-centric to cognition-centric alignment (Table~\ref{tab:paradigmshift}). 
Beyond merely cataloging surface-level intents, this framework models the underlying causal dependency graphs that constitute a user’s reasoning.  By capturing the concepts and causal links rather than isolated intents, the system can represent not only the structural rationale of user requirements (the ``why'') but also the cognitive motifs--reasoning pattern--that remain invariant across different domains. 
\begin{table}[h]
\centering
\small
\caption{Comparison between intent-centric and cognition-centric modeling.}
\begin{tabular}{lp{3cm}p{3cm}}
\textbf{Dimension} & \textbf{Intent-Centric} & \textbf{Cognition-Centric} \\
\hline
Modeling unit & Discrete Goals/intents & Concepts + causal links \\
\hline
Structure & Flat list or hierarchy & Causal dependency graph \\
\hline
Reasoning & Implicit & Explicit (via motifs) \\
\hline
Adaptation & Re-extract all intents & Propagate through dependencies \\
\hline
Reusability & Task-specific & Motifs transfer across tasks \\
\hline
Explainability & Surface-level ``What I want'' & Structural rationale ``Why I want it'' \\
\end{tabular}
\label{tab:paradigmshift}
\end{table}

    
    
\section{Challenges and Design Guidelines}
    \label{sec:3}

\subsection{Challenges in Human-LLM Cognitive Alignment}\label{sec:fs-challenges}

\emph{C1: Extracting Implicit Causal Rules from Natural Language}
Human reasoning is naturally organized through causal relationships~\cite{gopnik2004causal}, where decision-making depends on conditional logic, trade-offs, and contextual exceptions. Pearl's causal framework~\cite{pearl2009} provides a formal vocabulary for describing how one factor enables or constrains another; however, in human--LLM interaction, these structures are rarely stated explicitly. Instead, they emerge fragmentarily across turns through constraints and justifications. 
~

Intent-externalization systems~\cite{coscia_ongoal_2025,interQuest_2025_mei} identify \textit{what} users want, but often miss the \textit{causal rule} that explains when one consideration should override another. For example, when a user says \textit{``I usually prefer affordable options, but I'd pay more for verified reviews,''} existing systems often store two independent preferences rather than recognizing a conditional dependency.

\emph{C2: Maintaining Reasoning Structure Coherence Through Dynamic Revision}
User reasoning is inherently dynamic: prior assumptions are often qualified or refined rather than simply replaced. This mirrors the cognitive process of \textit{reasoning revision}~\cite{gardenfors1988}, where new utterances serve as evidence for updating a structured reasoning state through counterfactual reasoning~\cite{byrne2005} and probabilistic updating~\cite{griffiths2010}. 
~

Existing systems often treat turns independently~\cite{coscia_ongoal_2025} or accumulate preferences additively~\cite{interQuest_2025_mei}, while recent work on persistent AI memory shows that updating intent specifications themselves is a non-trivial interaction problem~\cite{vaithilingam_semanticcommit_2025}. This obscures how reasoning structures \emph{revise} rather than merely \emph{accrete}. If a user first says \textit{``I prefer budget hotels''} and later says \textit{``I'd pay more for verified cleanliness ratings,''} the issue is not that the earlier preference was wrong, but that it was context-bounded. A system that overwrites or naively accumulates such statements misrepresents the underlying reasoning structure.

\emph{C3: Enabling Cross-Task Reusable Reasoning Structure}
The value of extracting reasoning structures lies in their \textit{compositionality}---the ability to reuse modular schemas across related situations~\cite{tenenbaum2011,lake2015}. Once a reasoning pattern (e.g., \textit{``safety-first''}) is learned, users should not need to restate it from scratch in every new task.
~

TaskMind~\cite{yin_operation_2025} demonstrated cross-instance generalization within GUI tasks, and Memolet~\cite{yen_memolet_2024} showed the value of reifying reusable user-AI conversational memories, but conversational settings still lack mechanisms for assessing whether a reasoning pattern from one domain should transfer to another, and for bringing that transfer under explicit user review.
\subsection{Design Guidelines}\label{sec:design-goals}
Based on these challenges, we derive four design guidelines.

\textbf{DG1: Extract Candidate Concepts and Causal Dependencies Without Disrupting Flow} (\S\ref{sec:stage1}).  
To support causal reasoning, the system should extract concepts and dependency candidates as they emerge in dialogue, while deferring commitment until those candidates are grounded through interaction and runtime evidence.

\textbf{DG2: Represent Reusable Reasoning Patterns} (\S\ref{sec:stage2}).  
Because recurring reasoning often reappear across contexts, the system should abstract grounded local structures into reusable motifs that support later clarification, transfer, and cross-task reuse.

\textbf{DG3: Make Clarification, Revision, and Transfer Reviewable} (\S\ref{sec:stage3}).  
Reasoning structure revision should preserve structural continuity rather than overwrite prior reasoning. The system should therefore expose clarification, revision, and transfer as reviewable operations over a shared reasoning structure.

\textbf{DG4: Compile Reasoning into a Stable and Inspectable Interaction Substrate} (\S\ref{sec:stage3}--\S\ref{sec:system-interface}).  
Cognitive alignment must be legible and editable. The system should compile model-proposed and user-grounded structure into a stable visual substrate that users can inspect, question, and revise over time.

\section{CogInstrument: System Design} 
\label{sec:systemdesign}

The four design guidelines above converge on a core design principle: \textbf{represent user reasoning as a graph of causally linked concepts that can be composed into reusable motifs and revised in a principled, inspectable manner.}


\subsection{Concept Extraction and Causal Linking}
\label{sec:stage1}


\hl{\emph{Cognitive Concepts.}
We define a \textbf{concept} $c$ as the minimal semantic unit that participates in causal reasoning. Drawing from structured cognition literature, we distinguish four concept types: \emph{belief}, \emph{constraint}, \emph{preference}, and \emph{factual assertion}. 
These typed concepts serve as the atomic units of the cognitive graph and preserve semantic roles that are typically flattened in intent-extraction pipelines.} 

\hl{A \textbf{causal dependency} $c_i \xrightarrow{\phi} c_j$ indicates that concept $c_j$ is formed or influenced because of a reasoning process $\phi$ applied to $c_i$. Drawing on 
Pearl's causal framework} \cite{pearl2009} (Appendix \ref{apx:four_causal_operations}), we define a typed dependency schema with three relations: \textbf{enable}, which makes another concept actionable without determining its value; \textbf{constraint}, which restricts its feasible space; and \textbf{determine}, which directly specifies its value (detail definitions in Appendix~\ref{apx:causaltypes}; \ref{apx:cognitiveconcept_diagram}, Figure \ref{fig:concept_dependency}).

\emph{From Utterances to Concepts.}
Following this definition, the system extracts concepts $c$ from given a user utterance $u_t$ by identification, disambiguation, and validation. These candidates are not treated as committed state; they are later grounded through cross-turn evidence, function-call support, and structural consistency checks.

\emph{Causal Link Identification.}
Within these fixed schemas, LLM-assisted causal discovery proposes candidate dependency edges \cite{zecevic2023}. The runtime then grounds, questions, and stabilizes these candidates so that dependencies serve not only as inferential structure, but also as the editable backbone of the externalized reasoning graph. Although we illustrate the process with travel-style examples, the representation and interaction mechanisms are task-agnostic.

\begin{figure}[h]
    \centering
    \includegraphics[width=1.05\linewidth]{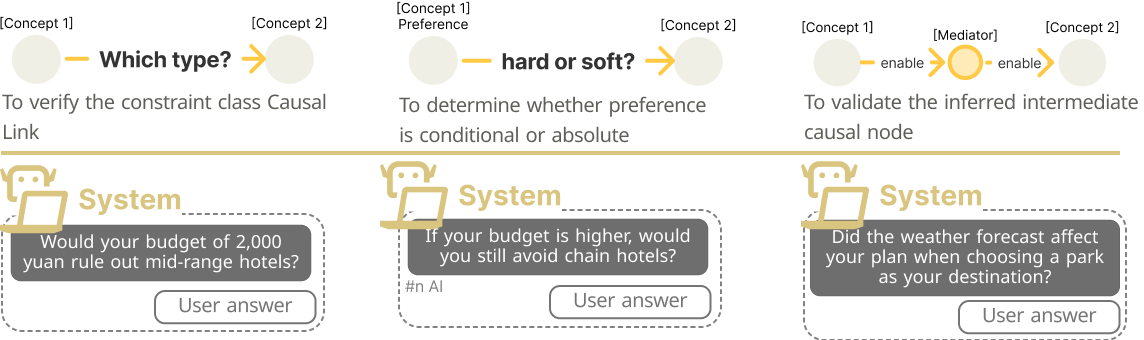}
    \caption{Representative clarification probes for different uncertainty patterns.}
    \label{fig:uncertainty_3probes}
\end{figure}
\emph{Validation Through Proactive Questioning.}
\label{sec:stage1_questioning}
When verification is necessary, the system probes high-value dependency candidates through three question types: \emph{direct confirmation}, \emph{counterfactual probing}, and \emph{mediation check}. User responses resolve ambiguity by confirming, weakening, or refining candidate dependencies (Figure \ref{fig:uncertainty_3probes}).

\subsection{From Causal Links to Cognitive Motifs}
\label{sec:stage2}

\begin{figure}[H]
    \centering
    \includegraphics[width=0.4\linewidth]{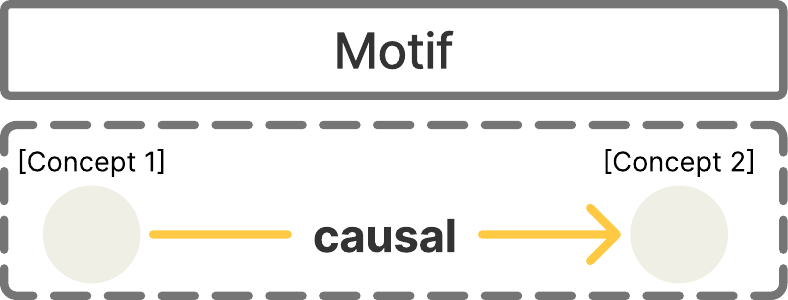}
    \caption{A motif is a reusable reasoning pattern with concepts and its dependency.}
    \label{fig:concept_motif}
\end{figure}
\emph{Cognitive Motifs.}
\hl{To support compositional reasoning, we introduce the notion of a \textbf{cognitive motif}: a \textbf{cognitive motif} $\mu$ is a reusable causal subgraph representing a common reasoning pattern} (Figure \ref{fig:concept_motif}). It is a minimal reasoning structure consisting of at least two concepts connected by at least one directed causal dependency.


Drawing from analogical reasoning \cite{gentner1983} and schema theory \cite{rumelhart1980}, we organize motifs into a small taxonomy of recurring reasoning patterns (Appendix~\ref{apx:cognitive_motif}, Table~\ref{tab:motifs_codebook}), including \emph{constraint}, \emph{preference}, \emph{trade-off}, \emph{sequential}, and \emph{conditional} motifs. 
Motifs function as reusable schemas that can be instantiated in one context and adapted in another while retaining explicit relational semantics. 

\emph{Motif Abstraction.}
With this taxonomy defining the motif vocabulary, the runtime instantiates motif candidates by matching grounded dependencies to a fixed set of reusable patterns. Formally, a cognitive motif is modeled as
\begin{equation}
\mu=(C_\mu,E_\mu,\phi_\mu),
\label{eq:motif}
\end{equation}
where $C_\mu\subseteq C$ is the set of concept nodes, $E_\mu\subseteq E$ the typed causal edges, and $\phi_\mu$ the abstract reasoning function (e.g., constraint propagation, trade-off resolution, or preference filtering). Validated motifs are stored as reusable patterns, retrieved across tasks as transfer candidates rather than copied answers, and composed around shared concepts to scale from local dependencies to task-level reasoning. These patterns are therefore not only internal representations, but also the basis for later review, revision, and transfer in interaction.

\subsection{Mixed-Initiative Revision, Clarification, and Runtime Architecture}\label{sec:stage3}

Once motifs are formed, the problem shifts from extraction to maintenance: the system must keep the reasoning structure inspectable and reusable without interrupting the user on every uncertain inference. \Cg{} therefore uses a mixed-initiative runtime in which local, low-impact updates are applied automatically, while clarification, non-local revision, and cross-task transfer are surfaced only when they are likely to change the active reasoning structure, the current plan draft, or future reuse (Appendix \ref{apx:systemdesign}, Figure \ref{fig:system}).

\textbf{State separation.}
\Cg{} maintains two distinct state layers: a \emph{cognitive state} and a \emph{task-plan state}. The cognitive state stores user-grounded concepts, task-specific motifs, clarification status, transfer candidates, task histories, and the reusable motif library. The task-plan state stores assistant-authored and co-authored planning artifacts for the current task, including drafts, comparisons, notes, and open questions with provenance labels such as assistant-proposed, user-confirmed, co-authored, and transfer-based. This separation preserves two key boundaries: motif patterns may be reused across tasks, but motif instances remain task-bound; likewise, assistant proposals may accumulate into a working plan immediately, but only user-confirmed evidence enters the cognitive layer.

\textbf{Selective clarification.}
\Cg{} surfaces clarification selectively rather than exhaustively. For a motif candidate $m$, the runtime computes
\[
I(m) = \alpha_u \, \mathrm{Unc}(m) + \alpha_s \, \mathrm{Cent}(m) + \alpha_c \, (1-\mathrm{Cov}(m)) + \alpha_t \, \mathrm{Risk}(m),
\]
where $\mathrm{Unc}(m)$ is motif uncertainty, $\mathrm{Cent}(m)$ structural centrality, $\mathrm{Cov}(m)$ evidence coverage, and $\mathrm{Risk}(m)$ transfer risk; $\alpha_u, \alpha_s, \alpha_c, \alpha_t$ are fixed coefficients tuned during pilots and frozen before the formal study, and $\tau$ is the clarification threshold. The system asks at most one explicit clarification question per turn, surfacing only the highest-impact candidate with $I(m) > \tau$ through lightweight confirmation, counterfactual probing, or mediation checks (\S4.1.2); lower-impact ambiguity remains provisional in the background. User responses update motif status and evidence attribution, for example activating a motif, weakening a dependency, or leaving a candidate pending.

\textbf{Structured revision.}
Revision is implemented as graph patching rather than hidden rewriting of conversation history. \Cg{} supports concept revision, confidence revision, and structure revision, and tracks motifs as \emph{active}, \emph{uncertain}, \emph{deprecated}, or \emph{cancelled}. Both user statements and system-side consistency checks can trigger candidate patches. The runtime computes a structural diff over affected concepts, dependencies, motifs, and downstream planning consequences: small local changes are committed automatically, while broader changes are surfaced for user review and approval. Importantly, assistant suggestions do not revise the user-grounded layer directly. They are written first to the task-plan state, and only explicit user confirmation promotes them into the cognitive layer.

\textbf{Cross-task transfer.}
Cross-task reuse operates over reasoning patterns rather than prior recommendations. At the start of a new task, \Cg{} retrieves candidate patterns from the motif library and instantiates them as transfer candidates. These candidates remain uncertain until supported by current-task evidence. Users may adopt, modify, or reject them, with clarification invoked only when the transfer remains materially ambiguous. In this way, transfer supports reuse without silently carrying over earlier conclusions.

\textbf{Grounding, stabilization, and implementation.}
Concept grounding and motif selection are handled by fixed runtime heuristics that fuse multi-turn evidence, downweight assistant-only support, and adapt thresholding to the local structural role of each candidate; all coefficients were tuned in pilots and frozen before the formal study. Each turn is compiled into a small graph patch over an acyclic backbone of \emph{enable}, \emph{constraint}, and \emph{determine} edges, while conflict edges remain explicit outside the backbone so tensions stay inspectable rather than being silently resolved. To keep the graph stable under ongoing revision, singleton-slot candidates are compacted, non-slot concepts are attached with A*-guided anchor selection \cite{hart1968} to preserve local continuity, and cyclic dependencies are detected with Tarjan's algorithm \cite{tarjan1972} and repaired by removing the weakest structural edge. For presentation, the repaired graph is laid out with a layered strategy inspired by Sugiyama et al.~\cite{sugiyama1981}, with position stabilization following Brandes and K\"opf \cite{brandes2002}, so recurrent structures remain visually comparable across turns. \Cg{} is implemented as a client--server web system. To support reproducibility under anonymous review, we provide in the supplementary materials an anonymized open-source implementation, a system architecture diagram, default parameter settings, and a compact summary of the runtime algorithms. These mechanisms are implemented as runtime heuristics rather than learned models or normative optimization objectives.

This runtime design gives the interface a clear division of labor: clarification handles consequential ambiguity, revision exposes non-local changes as inspectable patches, and transfer introduces reusable patterns as reviewable candidates rather than hidden carryover.

\subsection{Interaction Design}\label{sec:system-interface}

\begin{figure*}[t]
    \centering
    \includegraphics[width=\textwidth]{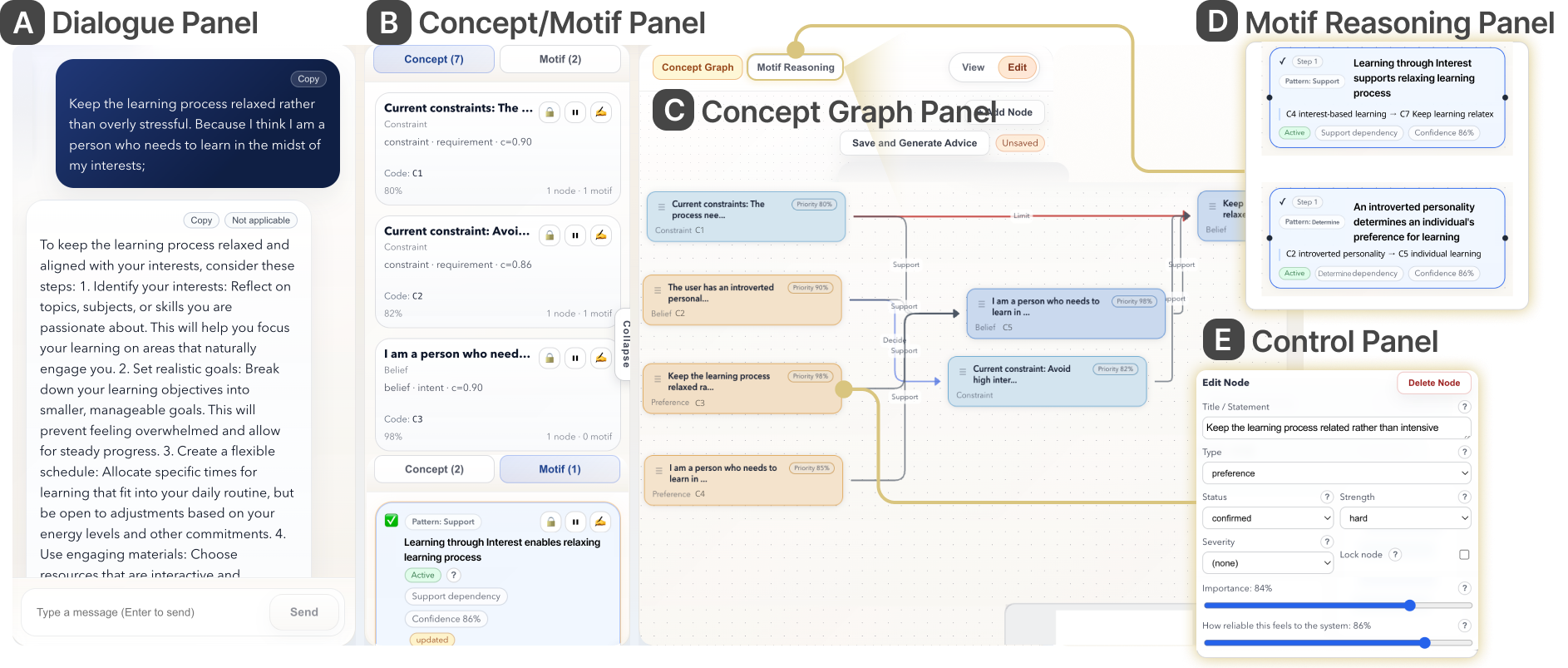}
    \caption{
    \Cg{} interface. Panels (A--E) provide synchronized views of the underlying reasoning state, ranging from high-level dialogue planning (A) and structural reasoning mapping (B--D) to direct intervention and patch management (E).
    }
    \Description{The \Cg{} interface includes a dialogue panel, a concept and motif list, a cognition graph, a motif reasoning panel, and a control panel for editing selected items.}
    \label{fig:interface}
\end{figure*}

\begin{figure}[h]
    \centering
    \includegraphics[width=0.8\linewidth]{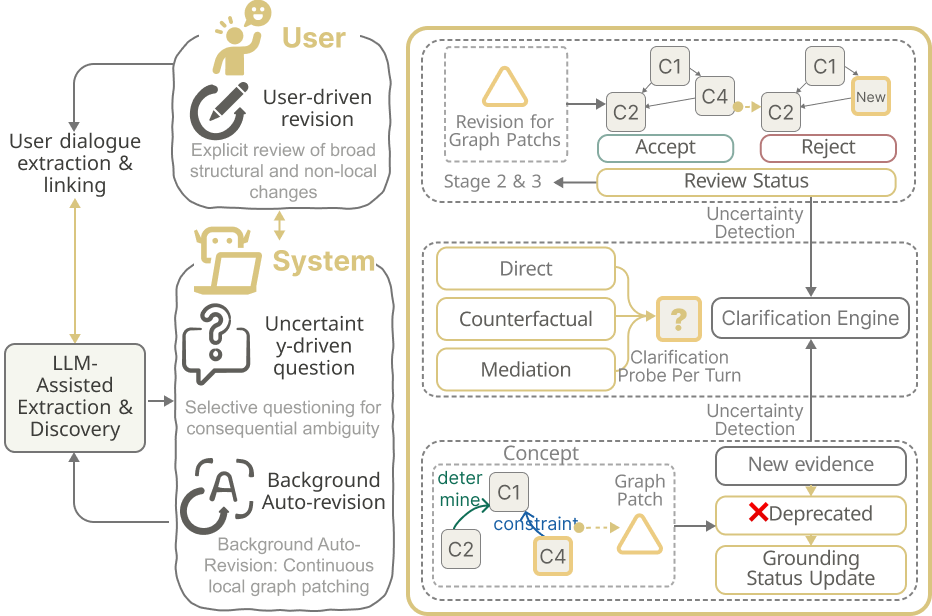}
    \caption{Three interaction modes: user-driven revision, uncertainty-driven question, and background auto-revision.}
    \label{fig:threeinteractionmodes}
\end{figure}

Figure~\ref{fig:interface} shows five coordinated panels over the same runtime state. The \textbf{A Dialogue Panel} is the primary channel for planning, clarification, and plan generation. \textbf{Panels~B--D} externalize the evolving reasoning structure at complementary levels: as editable lists of concepts, motifs, and transfer candidates; as a stabilized graph of typed dependencies; and as motif-level rationales, status, and local consequences. The \textbf{E Control Panel} provides targeted intervention, including direct edits and approval of surfaced patches. Rather than acting as separate modules, these views provide synchronized entry points into the same underlying reasoning object, allowing users to move between conversational interaction, structural inspection, and explicit correction without losing context. Figure~\ref{fig:threeinteractionmodes} summarizes how runtime operations become visible in interaction. We illustrate a user walkthrough to demonstrate \Cg{}'s cognitive alignment capability in Appendix \ref{apx:system-walkthrough}.


\section{User Study}
    To evaluate whether externalizing causal reasoning structure improves human--LLM alignment beyond intent capture, we conducted a within-subjects controlled study comparing our proposed \Cg{} with a baseline chat interface across two planning scenarios, each containing two sequential tasks
The study addressed two questions:
\begin{enumerate}
    \item[\textbf{RQ1}] How does \Cg{} externalize users' implicit causal reasoning and make its influence on model responses more inspectable?
    \item[\textbf{RQ2}] How does \Cg{} affect principled reasoning updates, cross-task reuse, and users' experience?
\end{enumerate}

\subsection{Participants}

We recruited 12 experienced users of LLM-assisted planning tools through purposive sampling from LLM-related online communities and university channels. This sampling strategy was appropriate because the study required participants who could make informed comparisons between a conventional chat workflow and \Cg{}'s reasoning-centered interaction. Eligibility required at least one year of LLM experience and at least weekly use. All participants completed a pre-study screening questionnaire on demographics and prior LLM use.
~
The final sample comprised 12 participants (4 male, 8 female; age 21--28) from research, software/engineering, creative practice, and student backgrounds (Appendix \ref{apx:participants}, Table~\ref{tab:participants}). Participants reported 2--5 years of LLM experience, and six reported daily LLM use.
~
All participants provided informed consent and received US\$14 for a 70--80 minute session. The study was approved by our institution's IRB.

\subsection{Scenarios and Tasks}

We designed two planning scenarios, each containing two sequential 10-minute tasks that shared underlying reasoning motifs while differing in surface domain. This allowed us to examine cross-task reuse while avoiding reliance on residual LLM conversation history.

\textbf{Scenario A (Travel Planning):} Task~A1 involved planning a trip to Hong Kong; Task~A2 involved planning a trip to Paris. The pair shared motifs such as budget--accommodation trade-offs and preferences for local experience.

\textbf{Scenario B (Personal Growth Planning):} Task~B1 involved designing a fitness plan; Task~B2 involved designing a study plan. The pair shared motifs such as procrastination management and time--energy trade-offs. Details are shown in Appendix \ref{apx:tasks}.

Each task had two phases: (1) \textbf{Open Planning}, in which participants explored the task freely, and (2) \textbf{Constraint Injection}, in which the researcher introduced three conflicting constraints mid-task to elicit revision and trade-off reasoning under changing conditions. Participants were asked to think aloud about the rationale behind their preferences and decisions, and at the end of each task they summarized their final plan as a reference artifact.

\subsection{Study Procedure}

\subsubsection{Procedure}

Sessions lasted approximately 70--80 minutes and were screen- and audio-recorded. Participants thought aloud throughout the session~\cite{Ericsson1980VerbalRA}. After a 10-minute introduction and consent process, each participant completed one condition (two tasks, 20 minutes), filled out the post-condition questionnaires, then completed the other condition (two tasks, 20 minutes), followed by the same questionnaires and a 15--20 minute semi-structured interview.

Conversation history was cleared between Task~1 and Task~2 in both conditions. This ensured that any cross-task carryover arose from participants' own reasoning or from \Cg{}'s transfer mechanism rather than from residual LLM context.
~
To counterbalance condition order and scenario assignment, participants were randomly assigned to two groups. Group~A ($N=6$) completed Baseline on Scenario~A and \Cg{} on Scenario~B; Group~B ($N=6$) completed the reverse. This yielded partial counterbalancing of condition order and scenario assignment while preserving the sequential task structure required to study transfer.

\subsubsection{Baseline System}

To isolate the contribution of cognitive externalization and alignment support, the baseline was implemented as a standard multi-turn chat interface using the same GPT-5.3 backend, token budget, system prompt, and task-specific RAG support as \Cg{}. It preserved full within-task conversation history and the same free-form conversational planning capability: participants could ask follow-up questions, add or revise constraints, and iteratively refine plans through ordinary dialogue as in \Cg{}. The baseline removed only the structure-centered mechanisms: no concept extraction, no graphical reasoning view, no editable dependencies or motifs, and no cross-task transfer. The comparison therefore varies the interaction paradigm rather than the underlying model or retrieval capability.

\subsubsection{Measures}
\label{sec:measures}

We collected post-condition questionnaires, think-aloud data, post-study interviews, and interaction logs.

\textit{Subjective Measures.}
After each condition, participants completed a study-specific 17-item questionnaire using 7-point Likert scales (Appendix \ref{apx:A1}) across six design-aligned constructs: \textit{Diagnosis Clarity}, \textit{Reasoning Externalization}, \textit{Dependency Grounding}, \textit{Revision Coherence}, \textit{Cross-Task Transfer}, and \textit{Trust \& Control}. These study-specific constructs served as the primary subjective outcomes. Participants also completed the System Usability Scale (SUS) and Raw NASA-TLX as secondary generic measures. After both conditions, participants took part in a 15--20 minute semi-structured interview probing reasoning capture, structure inspection, revision, transfer, and perceived control (Appendix~\ref{apx:interviewguide}).

\textit{Interaction Logs.}
We collected time-stamped raw interaction logs in both conditions. In both systems, logs included user submissions, system responses, and task timestamps. In \Cg{}, logs additionally captured concept edits, motif edits, edge edits, and transfer-uptake actions as explicit interface events.

For behavioral comparison, we derived coded event timelines from the raw logs and aligned dialogue turns. Two researchers coded four textual event types observable in both conditions: text restatement, text correction, text new/rewrite, and text-based cross-task reuse. These codes were used descriptively to characterize revision pathways rather than to infer latent cognition. Structural and transfer events in \Cg{} were identified directly from instrumented system logs, because these actions had explicit event records at the interface layer, and were aligned to the same timelines. The resulting timelines were used to examine shifts between text-dominant and structure-level interaction, revision pathways, and transfer uptake over time. Coding disagreements were resolved through discussion to consensus.

\subsubsection{Data Analysis}
We employed a mixed-methods approach to address our research questions. RQ1 (Diagnosis, Externalization, Grounding) and RQ2 (Revision, Transfer, Trust, Control) were evaluated by triangulating questionnaire constructs, interaction logs, and qualitative accounts. For quantitative data, construct scores were calculated by averaging corresponding questionnaire items. Given the within-subject design and ordinal nature of the data, we used paired Wilcoxon signed-rank tests with matched-pairs rank-biserial correlations ($r_{rb}$) as effect sizes. This procedure was applied to aggregate constructs, individual items (as exploratory follow-ups), and secondary measures (SUS, NASA-TLX). 
Because the 17-item questionnaire was designed for this study, the six construct summaries are treated as design-aligned aggregates rather than evidence of latent-factor structure. Internal consistency was assessed descriptively; full item wording, item-to-construct mapping, and reliability estimates are reported in the appendix. 

Interaction-log data were analyzed descriptively to characterize revision pathways, structural interaction, and transfer uptake across conditions. Concurrent think-aloud transcripts and post-study interviews were analyzed by two researchers using Reflexive Thematic Analysis (RTA)~\cite{braun2019reflecting}, with iterative reading, memoing, discussion, and theme development. Findings are reported as integrated analytic claims organized by the research questions, bringing questionnaire results, log patterns, qualitative themes, and representative participant quotes into correspondence.

\section{Findings}
    \label{sec:findings}

\begin{figure}[t]
  \centering
  \includegraphics[width=\linewidth]{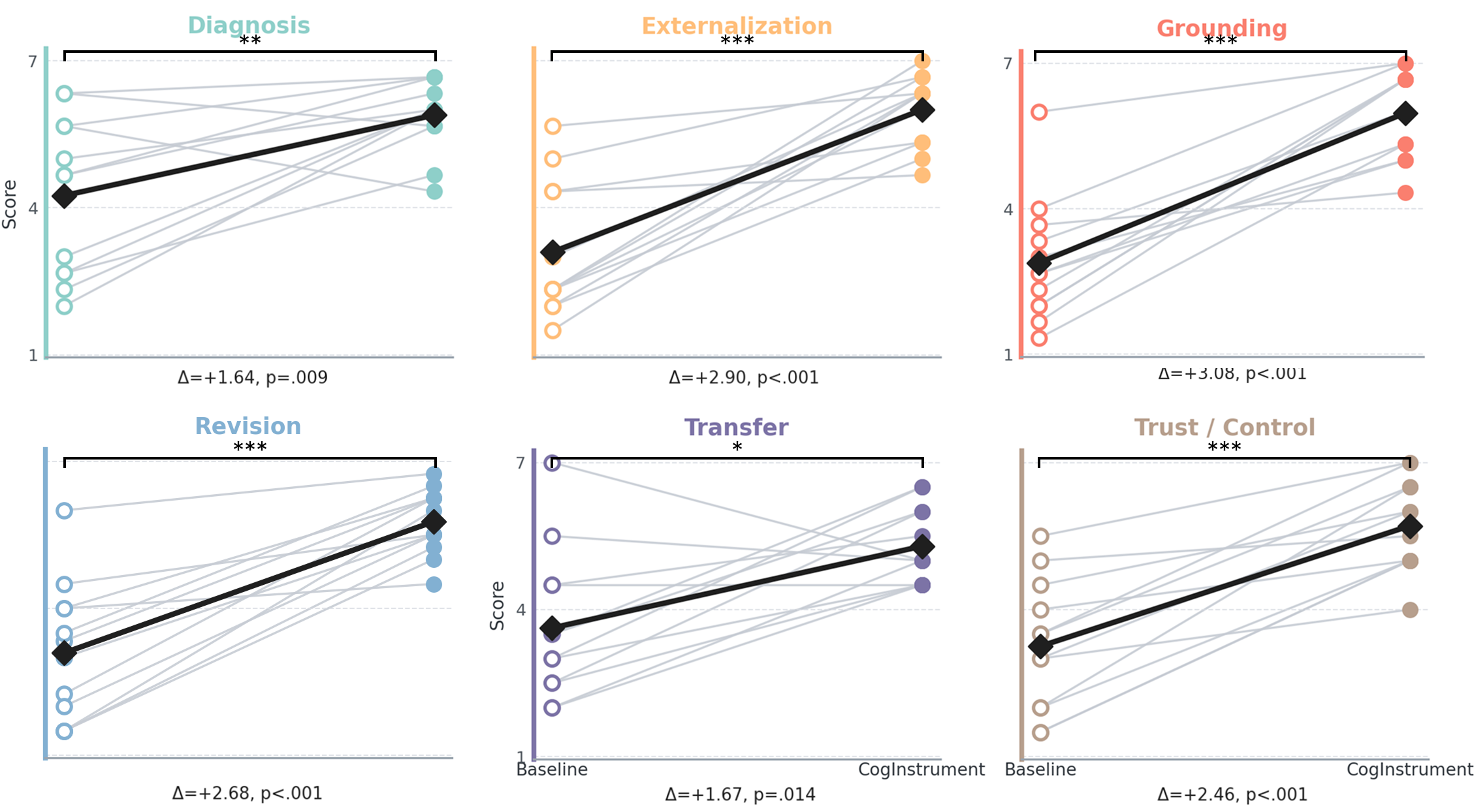}
  \caption{Paired participant trajectories and condition means for the six grouped Likert constructs. 
  }
  \label{fig:likert-constructs}
\end{figure}

Figure~\ref{fig:likert-constructs} shows a consistent shift in favor of \Cg{} across the six grouped constructs. At the overall level, the 17-item score increased from $M=3.34$ in Baseline to $M=5.80$ in \Cg{} ($p<.001$), indicating a broad improvement beyond ordinary prompt exchange.
~
All six construct summaries favored \Cg{} (all $p \le .014$). The strongest advantages appeared in \textit{Dependency Grounding} ($2.89 \rightarrow 5.97$, $p<.001$), \textit{Reasoning Externalization} ($3.10 \rightarrow 6.00$, $p<.001$), and \textit{Revision Coherence} ($3.09 \rightarrow 5.77$, $p<.001$), followed by \textit{Trust \& Control} ($3.25 \rightarrow 5.71$, $p<.001$), \textit{Cross-Task Transfer} ($3.62 \rightarrow 5.29$, $p=.014$), and \textit{Diagnosis Clarity} ($4.25 \rightarrow 5.89$, $p=.009$). Taken together, these results suggest that participants did not merely prefer the system overall; they experienced it as making their reasoning easier to externalize and inspect, the model's dependencies easier to understand, revisions more coherent, transfer more usable, and interaction more controllable.

Secondary measures did not indicate a generic usability advantage (SUS: $p=1.000$; Raw NASA-TLX composite: $p=.507$), although \textit{Mental Demand} was higher under \Cg{} ($p=.041$), suggesting that the system shifted effort toward more deliberate reasoning and revision rather than reducing workload overall.


\subsection{System-Aligned User Reasoning and Model Response (RQ1)}
\label{sec:finding_alignment}
\subsubsection{Externalized Structure Improved Self-Reflected Reasoning (DG1)}
Participants reported subjectively better support for diagnosing output mismatches (Q1: $Mdn_{Cg}=6.00$, $Mdn_{B}=5.00$, $p=.008$, $r=.526$), distinguishing AI misunderstanding from their own underspecification (Q2: $Mdn_{Cg}=6.50$, $Mdn_{B}=4.00$, $p=.016$, $r=.479$), and locating an entry point for correction (Q3: $Mdn_{Cg}=6.00$, $Mdn_{B}=4.50$, $p=.021$, $r=.665$). Interview accounts suggest that this diagnostic benefit came from making dependencies and conflicts visible rather than leaving them embedded in linear dialogue. P7 described the graph as a form of cognitive ``typesetting,'' while P11 noted that it could \textit{``intuitively display contradictions between user requirements, such as the tension between a limited budget and high consumption desires,''} which would otherwise remain hard to isolate in ordinary chat.

The graph also strengthened participants' ability to express the reasoning behind decisions (Q4: $Mdn_{Cg}=6.00$, $Mdn_{B}=3.00$, $p=.003$, $r=.84$), see dependencies among ideas (Q5: $Mdn_{Cg}=6.50$, $Mdn_{B}=3.00$, $p=.002$, $r=.88$), and notice their own thinking process (Q6: $Mdn_{Cg}=6.00$, $Mdn_{B}=3.50$, $p=.002$, $r=.87$). P2 described this as a ``moment of recognition,'' where the system surfaced an implicit preference---such as the historical interest behind a location choice---that had previously remained tacit. P1 characterized the graph as a ``mirror'' of the overall context of their statements, while P3 said the canvas ``not only aligned with my thinking, it clarified and even strengthened it.'' 
Participants also described the graph as prompting them to articulate the \textit{why} behind their goals. P3 noted that the system encouraged them to move from an unstructured preference such as wanting to visit Notre Dame to a more explicit rationale grounded in historical interest. 

The logs reinforce this interpretation. In several participants' timelines, interaction began with natural-language grounding and shifted toward concept- or edge-level intervention only after an initial scaffold had become visible. This suggests that the graph did not merely record what participants had already said; it changed how they diagnosed both the task and their own reasoning. Even in text-dominant traces, the externalized structure functioned as a scaffold for multi-turn reasoning by making prior assumptions visible enough to inspect rather than repeatedly restate. Together, these findings suggest that the graph served as a shared diagnostic surface rather than a passive record of prior dialogue.

\subsubsection{Causal Transparency Made Alignment Inspectable and Steerable (DG1)}
Participants reported that the system not only understood what they preferred, but also why those preferences mattered (Q7: $Mdn_{Cg}=6.00$, $Mdn_{B}=3.00$, $p<.001$, $r=.636$), which made responses feel more grounded in their reasoning (Q8: $Mdn_{Cg}=6.00$, $Mdn_{B}=2.50$, $p<.001$, $r=.627$). This perceived grounding was repeatedly tied to visibility into how constraints propagated through the reasoning structure rather than to output fluency alone.

By transforming reasoning into nodes and edges with explicit strengths and relations, system provided a more inspectable hierarchy for tasks. P9 likened this structured input to a physical control mechanism: \textit{``I feel this product is like giving me a `rein to steer' the large model. When these conditions are placed in the same graph and even their importance is visible, I can clearly know whether the AI is actually focusing on the points I care about.''} P6 similarly emphasized the auditing value of the graph: \textit{``Without motifs, you have no way of knowing how the AI assigns weights within your statement. But through motif visualization, you can clearly see what its decisions are based on and better control its direction.''} 

The logs clarify why this mattered. Even participants who rarely manipulated the graph directly still returned to it as an inspectable control representation when assessing whether the model had drifted from their priorities. In this sense, alignment was supported not only by editable structure, but by visible structure that made deviations diagnostically legible and therefore steerable.

\subsection{\Cg{} Facilitated Experience and Interaction with Reasoning Structure (RQ2)}
\subsubsection{Reviewable Revision Supported Principled and Traceable Reasoning Updates (DG2)}
\label{sec:finding_revisible}

\textbf{Structural Editing Let Users Revise Relationships, Not Just Replace Content}
The system shifted revision from \textit{whole-turn text overwriting} to \textit{structural reconfiguration}. When participants changed their minds, they reported stronger support for updating reasoning in a structured manner (Q9: $Mdn_{Cg}=6.00$, $Mdn_{B}=2.50$, $p<.001$, $r=.652$) and for considering how local changes would affect related decisions and downstream implications (Q11: $Mdn_{Cg}=6.00$, $Mdn_{B}=3.00$, $p<.001$, $r=.636$). P3 demonstrated this by resolving a logical conflict not by restating preferences, but by editing a causal edge: \textit{``After I changed the relationship from `constraint' to `support,' the system understood my intent better.''} 
\hl{P10 similarly noted ``re-edit motifs'' helped align system-generated structures with users' own internal vocabulary, keeping revision personally meaningful rather than merely machine-readable.}

The logs align with this interpretation. One recurrent pattern was a progression from early text-based grounding to later structure-level maintenance: once a workable scaffold had formed, participants shifted from broad natural-language restatement to more localized concept or edge intervention. This indicates that reviewable revision was not only experienced subjectively as more principled; it also changed the unit of interaction from whole-turn correction to local structural repair. 

\textbf{Targeted Edits Reduced Repair Cycles and Interruption Cost}
Motif-based structure provided a tangible advantage by localizing alignment errors. When the LLM failed to capture an intent correctly, participants reported stronger support for identifying which earlier judgments needed to be revisited after a change (Q12: $Mdn_{Cg}=5.50$, $Mdn_{B}=3.50$, $p=.003$, $r=.561$) and could often repair a misunderstanding by editing a specific concept, motif, or edge rather than re-describing the entire requirement from scratch. P5 highlighted this efficiency: \textit{``If I relied solely on text, I'd have to interrupt the AI every time. But with the motif format, I can be very explicit: this specific node is wrong. I only focus on fixing that one part without back-and-forth corrections.''} P6 described a similar case in which a misunderstood learning preference could be corrected directly in the graph. 
\hl{P2 further noted that the graph made it possible to identify a ``core bottleneck node'' that was disproportionately shaping a complex trip plan.} 

Across participants, this targeted repair stood in stark contrast to baseline conversational correction, where fixing one misunderstanding often risked losing context or re-introducing previous errors. The log timelines make this contrast behaviorally visible: explicit concept-, motif-, and edge-level edits appeared only in \Cg{}, and these operations were often interleaved with rather than replaced by textual interaction. Participants used structural edits when they needed precision, not as a blanket substitute for language.

\textbf{Iterative Graph Updates Created a Traceable Record of Evolving Reasoning}
The ability to observe how the graph changed across turns was described as distinct from and more useful than reviewing conversation history. Participants reported stronger support for tracking how decisions evolved over time (Q10: $Mdn_{Cg}=6.00$, $Mdn_{B}=3.00$, $p=.004$, $r=.552$), and interview accounts framed the canvas as a persistent audit trail rather than a transient visualization. P6 emphasized this directly: \textit{``Every intention I conveyed throughout the entire conversation was recorded in the graph and could be viewed at any time; therefore, it was fully traceable.''}


The logs help explain why this traceability mattered. Rather than appearing as hidden rewrites in conversation history, revisions accumulated as visible sequences of local updates over time. This made changing one's mind feel less like restarting the interaction and more like maintaining an evolving reasoning artifact.

\subsubsection{Motif Transfer Across Tasks Reduced Initiation Effort but Required Scope Negotiation (DG3)}
\label{sec:finding_compositional}

Compared with the larger gains for externalization and revision, the quantitative gain for transfer was more modest, suggesting that reuse was valuable but not automatic.

\textbf{Abstract Motifs Supported Cross-Domain Adaptation}
Participants reported that prior reasoning helped them start a new task more effectively (Q13: $Mdn_{Cg}=5.00$, $Mdn_{B}=4.00$, $p=.023$, $r=.456$). Unlike ordinary LLM memory, which often operates as opaque carryover, \Cg{} surfaced reusable motifs as explicit candidates for reuse. P10 captured this distinction succinctly: \textit{``It provides transferable motifs for me to choose from, whereas ordinary systems are just `context memory'.''}

Interview accounts suggest that what transferred was not surface content but reusable structure. P1 described a ``limited time'' constraint motif that remained useful across domains, while P3 identified higher-order patterns such as conflict resolution and priority balancing that transferred in a non-surface manner. The log data qualify this result: transfer uptake was selective rather than routine, and when it appeared, it was more often concentrated in Task~2 or after substantial current-task grounding had already occurred. This suggests that reusable reasoning patterns became actionable only once participants had enough current-task structure to judge what truly generalized.

\textbf{Scope Negotiation Kept Transfer Under User Control}
Participants also reported that prior reasoning transferred more naturally when they could actively judge its relevance (Q14: $Mdn_{Cg}=5.00$, $Mdn_{B}=3.00$, $p=.025$, $r=.462$). Rather than passively accepting transfer suggestions, they engaged in \textit{scope negotiation} to decide which prior constraints or patterns should survive in the new context. P11 noted that this controllability was especially important because commercial models typically function as black boxes: \textit{``Commercial models are black boxes... this system is different; that controllability is important, especially when defining new tasks.''}

This mechanism allowed users to refine logic while inheriting structure. P7 observed that selectively retaining earlier constraints prevented planning from ``drifting'' because of obsolete conditions. P11 further described accepting a recommended structure while making local modifications to fit a new task, such as migrating from a fitness plan to a study plan. The logs reinforce this point by showing that transfer was typically intertwined with later concept or motif updates rather than executed as one-shot reuse. In practice, transfer functioned as a reviewable starting point rather than as hidden memory.

\subsubsection{Externalizing Reasoning Strengthened Trust and Sense of Control (DG4)}
\label{sec:finding_trust}

\textbf{Trust Arose from Structural Legibility Rather Than Output Quality Alone}
Participants reported stronger trust in the system's reasoning representation (Q15: $Mdn_{Cg}=5.50$, $Mdn_{B}=2.50$, $p<.001$, $r=.633$), and interview accounts suggest that this trust came less from output polish than from being able to inspect the reasoning process that produced an answer. By externalizing constraints into nodes, relations, and strengths, \Cg{} turned the model's reasoning into a legible object for human verification.

Participants also described trust as something that could be repaired through structural legibility. P6 explained: \textit{``When it misunderstands me, my trust drops. But because it lists every reason and every step of the thinking process, I can modify it with a strong sense of control. This process not only restores my trust but even further enhances it.''} The logs help interpret this result: several participants used the graph less as a frequent editing surface than as an externalized reference for checking whether the model's current reasoning remained aligned. Trust therefore emerged from inspectability as much as from direct manipulation.

\textbf{Higher Entry Cost Yielded Stronger Control and More Proactive Steering}
Although \Cg{} introduced a higher cognitive cost, participants still reported stronger control over how reasoning was captured (Q16: $Mdn_{Cg}=6.00$, $Mdn_{B}=4.00$, $p=.002$, $r=.580$) and a stronger sense that the interaction went beyond merely stating goals (Q17: $Mdn_{Cg}=6.00$, $Mdn_{B}=2.50$, $p<.001$, $r=.606$). P12 summarized this as a ``high-threshold, high-reward'' experience: \textit{``The entry barrier is indeed higher... but once mastered, it is significantly superior to traditional `black-box' interactions in terms of interactive effect and the sense of manipulation.''} P5 similarly linked trust to intellectual ownership of the reasoning structure: \textit{``It helped me list priorities and organize which dimension each element belongs to---it makes me feel like I am the owner of this logic.''}

This interpretation is consistent with the secondary measures above: \Cg{} increased \textit{Mental Demand} without reducing overall usability or increasing overall workload. Rather than making planning generically easier, the system asked users to engage more explicitly with reasoning structure and rewarded that effort with stronger control. P1 captured the resulting shift in agency: \textit{``The interaction becomes directional. When you have the logic graph and reasoning feedback, you know exactly what adjustments to make next. This makes the answers more accurate because you are clearer about what you truly want.''}

The logs reveal that this stronger agency did not depend on everyone becoming a heavy graph editor. Some participants remained strongly text-dominant across both tasks, yet still described the graph as giving them ``a rein on the LLM.'' This is important for interpreting the system's value: the benefit of cognition externalization lay not only in direct structural manipulation, but also in making the model's evolving reasoning visible enough to steer, contest, and selectively reuse.

\section{Discussion}
    
Rather than simply demonstrating that structural externalization is possible, our findings reveal specific mechanisms by which \Cg{} alters human--LLM interaction. 


\subsection{\textbf{Mechanisms of Cognition-Centric Alignment}} 

\emph{Externalized Reasoning as a Shared Coordination Object}
The primary contribution of the motif graph is not that it perfectly maps human cognition, but that it establishes a shared coordination object for negotiation. 
Prior systems that externalize either user-side goals or model-side interpretations \cite{coscia_ongoal_2025,gmeiner_intent_2025,riche_ai-instruments_2025,zhang_neurosync_2025,yin_operation_2025} often leave the LLM and the user operating over separate representations. 
In contrast, \Cg{} utilizes the human reasoning as a boundary object that anchors the LLM's output to the user's actual reasoning, which is consistent with the vision of GUMs \cite{shaikh_creating_2025} materializing a user's knowledge and beliefs as a shared computational representation \cite{yang2018grounding}. However, while GUMs aggregate broad, cross-domain context from unstructured observations.  
\Cg{} targets the \textbf{micro-level reconfiguration of causal structure within reasoning tasks}. As reported (\S\ref{sec:finding_alignment}), the visual graph structure is used to to verify that underlying logic has been registered, the graph serves as a boundary object that anchors the LLM's output to the user's actual reasoning (\S\ref{sec:finding_alignment}, \S\ref{sec:finding_trust}). 
This shifts alignment from merely tracking end-state goals (OnGoal \cite{coscia_ongoal_2025}, NeuroSync \cite{zhang_neurosync_2025}) to maintaining a shared, reviewable reasoning substrate—allowing users to exercise direct control over the both human and model's ``internal understanding'' to maintain high-agency collaboration. 

\emph{Reviewability over Automation}
A central mechanism driving users' agency is the system's emphasis on reviewability. 
As highlighted by revision and cross-task transfercoherence (\S\ref{sec:finding_revisible}, \S\ref{sec:finding_compositional}), the system’s efficacy stems from treating belief updates as reviewable operations rather than opaque backend state changes. Consistent with \textit{SemanticCommit}’s approach to detecting semantic conflicts in large-scale intent specifications via knowledge graphs\cite{vaithilingam_semantic_2025}, our system posits that reviewability is fundamental to human-AI grounding \cite{clark1991grounding,shaikh2025navigating}. However, while \textit{SemanticCommit} focuses on macro-level information conflicts and versioning, our system delves into the micro-level reconfiguration of concept dependencies during reasoning. Users valued the ability to perform localized repairs on causal links without disrupting the entire task context (\S \ref{sec:finding_revisible}). This suggests that in complex reasoning tasks, designing for inspectable, mixed-initiative intervention \cite{bradshaw2003dimensions,horvitz1999principles} is more effective for alignment than attempting to achieve zero-shot extraction perfection \cite{heer2019agency}.

\emph{Reusability and Heterogeneous Appropriation of Mental Models.}
Introducing a structural intermediate layer inevitably alters interaction cost, yet our study reveals a productive trade-off: users reported higher mental demand (\S\ref{sec:findings}), but overall usability and satisfaction increased (\S\ref{sec:finding_trust}). We interpret this through the lens of \emph{instrumental genesis}~\cite{rabardel2002people,beaudouin2000instrumental}, 
which distinguishes an artifact's designed functions from the \emph{schemes of use} that individuals construct around it. \Cg{} was not uniformly appropriated: some users treated the graph as an \emph{inspectable coordination surface}, scanning dependency structure before committing to edits; others engaged in \emph{localized structural repair}, intervening 
at specific nodes once a mismatch was identified; a smaller subset used 
the graph as an iterative composition scratchpad, building on prior edits 
across tasks. That these strategies remain viable is precisely because the 
system preserves \emph{composability}: prior edits accumulate as 
coordination anchors rather than being discarded between turns. Crucially, 
the elevated mental demand this produces is not friction but 
\emph{germane load}~\cite{Sweller1988}---effort invested in constructing 
durable mental models of causal task structure, rather than in 
trial-and-error prompt reformulation. The diversity of appropriation 
strategies is thus evidence that users are engaging in genuine 
schema-building, not merely tolerating a more complex interface. 

\subsection{\textbf{Classic Graph Algorithms as Interaction Infrastructure}}
\label{sec:discussion-graph-infra}
Developing and testing \Cg{} was itself an iterative Research through Design process \cite{zimmerman2007rtd}. Across iterations, we found a recurring tension: although LLMs are strong at semantic interpretation, the structures they suggest are often too unstable to serve directly as interaction objects. Even with constrained extraction and staged prompting, users still had to anticipate system behavior while specifying goals \cite{subramonyam_bridging_2024,wu_ai_2022}. Because fluent causal language does not guarantee coherent causal structure, model-generated graphs can drift, duplicate, or cycle across turns \cite{zecevic2023}. We therefore treat structural reliability as a system guarantee: the model proposes concepts and links, while a deterministic compiler enforces checkable invariants over a DAG backbone, including sparsity, connectivity, traceable revision, and tension preservation \cite{pearl2009,gardenfors1988}. Although human reasoning is not always DAG-like, supporting richer non-DAG structures would substantially increase modeling and interaction complexity. We therefore adopt a DAG backbone as a pragmatic substrate for the current system and leave richer structural modeling to future work.

\subsection{Limitations and Future Works}

While \Cg{} demonstrates the efficacy of externalizing causal reasoning for human--LLM alignment, several limitations reflect the boundaries of our current evidence and offer opportunities for future research.
~
First, our evaluation relies primarily on subjective, self-reported measures derived from a study-specific questionnaire aggregate rather than validated latent scales. Crucially, we lack an objective \textit{plan-quality metric} to evaluate the final artifacts, and we do not have a ground-truth dataset to quantitatively measure the extraction fidelity (e.g., precision and recall) of the generated concepts and causal dependencies. Furthermore, our small sample size ($N=12$) consisted of experienced LLM users; novice users might exhibit different behavioral patterns and interaction bottlenecks.

Second, the system’s performance is intrinsically tied to the reasoning capabilities of the underlying LLM, and its robustness across different model architectures remains unexplored. Additionally, \Cg{} currently focuses on a single granularity of reasoning. In highly complex scenarios, users may require multi-level goal tracking, where local operational constraints are hierarchically nested within global strategic motifs. 

Third, our study was strictly situated within the domain of planning tasks. The generalizability of motif-based externalization to other contexts—such as rapid decision-making or fluid brainstorming—remains unverified. In tasks with lower cognitive load or higher temporal pressure, the overhead of structural manipulation might outweigh its benefits, leading to ``representation fatigue.''

Finally, \Cg{} introduces a steeper learning curve compared to traditional ``black-box'' chat interfaces. As noted by participants (e.g., P12), this entry barrier can make initial interactions daunting. Moreover, the positive reception may be partially influenced by a novelty effect. Future longitudinal studies are needed to observe how users' mental models evolve over weeks, and whether extended use fosters genuine co-reasoning or an over-reliance on the system's externalized structures.

\balance
\section{Conclusion}


This paper introduces \Cg{}, a framework that bridges the human--LLM cognitive gap by transforming implicit reasoning into a manipulable graph of \textit{motifs} and \textit{causal dependencies}. By shifting from opaque prompting to structured representation, \Cg{} turns tacit logic into a shared boundary object for human--AI negotiation.
~
Our study demonstrates that \Cg{} enhances coordination through the \textbf{externalization} of causal reasoning, the \textbf{reviewability} of model updates, and the \textbf{selective transfer} of reasoning patterns. Participants transitioned from "prompting and praying" to proactive guidance as the system made underlying logic inspectable and editable. Ultimately, by treating causal dependencies as first-class interaction primitives, \Cg{} provides a foundation for more transparent, predictable, and steerable AI interfaces.

\bibliographystyle{ACM-Reference-Format}
\bibliography{CogInstruments,primo}

\appendix
\definecolor{boxbg}{RGB}{248,248,248}    
\definecolor{boxborder}{RGB}{245,245,245} 








\section{Framework: Representative Model}
\subsection{Types of Causal Relations in Reasoning}
\label{apx:four_causal_operations}
\begin{enumerate}
    \item \textbf{Direct causation}: $A \to B$ (``Increasing budget enables higher-quality hotels'')
    \item \textbf{Mediated causation}: $A \to M \to B$ (``Travel dates $\to$ seasonal pricing $\to$ total cost'')
    \item \textbf{Confounding}: $A \leftarrow C \to B$ (``Family size affects both transportation needs and accommodation type'')
    \item \textbf{Intervention}: $do(A)$ (``What if we change the destination city?'')
\end{enumerate}

\subsection{Causal Link Identification} \label{apx:causaltypes}
\begin{itemize}
    \item \textbf{Enable $\times$ Direct / Mediated causation}: Enabling dependencies are often supported by direct or mediated causal reasoning, where users infer how a change in one concept propagates feasibility to downstream options.
    \item \textbf{Constraint $\times$ Confounding}: Confounding structures often function as cognitive constraints, as a single latent factor narrows multiple decision dimensions simultaneously.
    \item \textbf{Determine $\times$ Intervention (do-operator)}: Intervention reasoning frequently precedes determination, as users mentally simulate alternative commitments before locking in a decision.
\end{itemize}

\subsection{Causal Link Types among Concepts} \label{apx:cognitiveconcept_diagram}
\begin{figure}[H]
    \centering
    \includegraphics[width=\columnwidth]{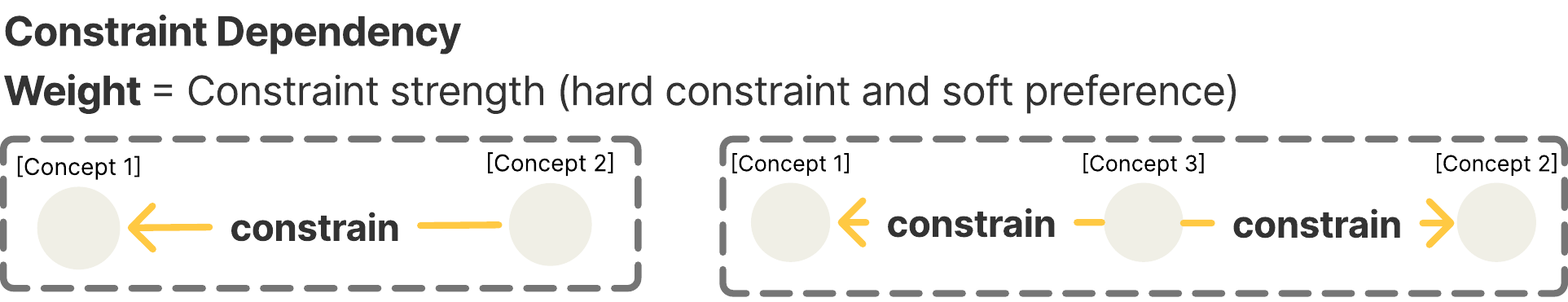}\hfill
    \includegraphics[width=\columnwidth]{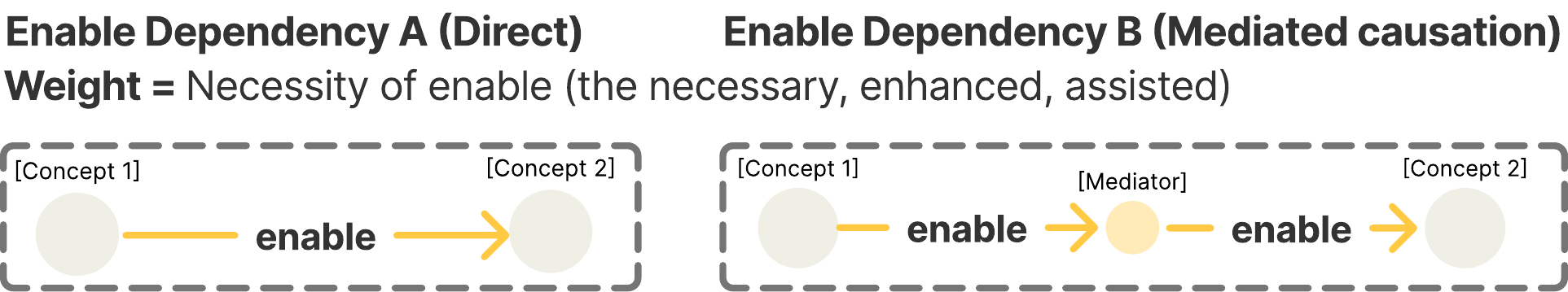}\\[0.25em]
    \includegraphics[width=0.5\columnwidth]{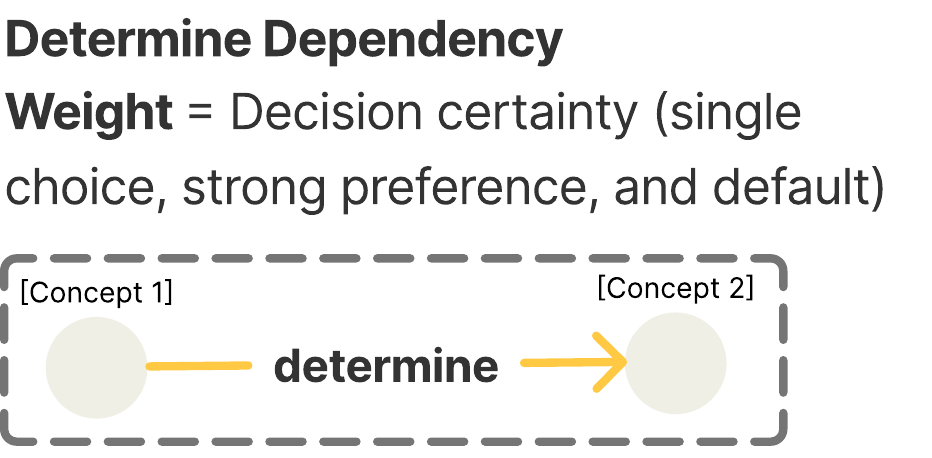}
    \caption{\Cg{} consists of three dependency types among concepts: enable, constraint, and determine. A motif is a reusable cognitive dependency pattern including at least two concepts.}
    \Description{Three types of causal structure among cognitive concepts: enable, constraint, and determine. A cognitive motif is a reusable cognitive dependency pattern containing at least two concepts and causal structure between them.}
    \label{fig:concept_dependency}
\end{figure}


\subsection{Cognitive Motif}\label{apx:cognitive_motif}

\paragraph{Formula Definition}
A cognitive motif is formally:
\begin{equation}
\mu = (C_\mu, E_\mu, \phi_\mu)
\end{equation}
where:
\begin{itemize}
    \item $C_\mu \subseteq \mathcal{C}$: concept nodes in the motif
    \item $E_\mu$: causal edges within the motif
    \item $\phi_\mu$: abstract reasoning function (e.g., ``constraint propagation'')
\end{itemize}

\emph{Example: Budget Constraint Motif}

\begin{lstlisting}
Motif: BUDGET_CONSTRAINT
Concepts: {budget_limit, option_cost, option_feasibility}
Causal Structure:
  budget_limit --[compare]--> option_cost
  option_cost --[filter]--> option_feasibility
Generalization:
  Applies to: hotels, restaurants, activities,
              transportation, shopping
\end{lstlisting}
\begin{table}[H]
\centering
\small
\begin{tabular}{@{}p{2.8cm}p{2.8cm}p{2.8cm}@{}}
\toprule
\textbf{Motif Name} & \textbf{Pattern} & \textbf{Example Instantiation} \\
\midrule
\texttt{budget-constraint} & Decision constrained by cumulative cost & ``Can't book \$200 hotel if \$2800 already spent'' \\
\addlinespace
\texttt{weather-adaptation} & Location/activity choice conditioned on forecast & ``Indoor museum if raining, park if sunny'' \\
\addlinespace
\texttt{temporal-precedence} & Task B requires completion of Task A & ``Book hotel before planning daily activities'' \\
\addlinespace
\texttt{preference-filtering} & Options filtered by stated preferences & ``Kid-friendly filter for family travelers'' \\
\addlinespace
\texttt{comparative-selection} & Choose best option based on criteria & ``Select restaurant with highest rating under budget'' \\
\bottomrule
\end{tabular}
\caption{Core cognitive motifs in the Operation-Cognition Knowledge Base}
\label{tab:motifs_codebook}
\end{table}






\subsection{System Design}\label{apx:systemdesign}
\begin{figure*}[htbp]
    \centering
    \includegraphics[width=0.9\linewidth]{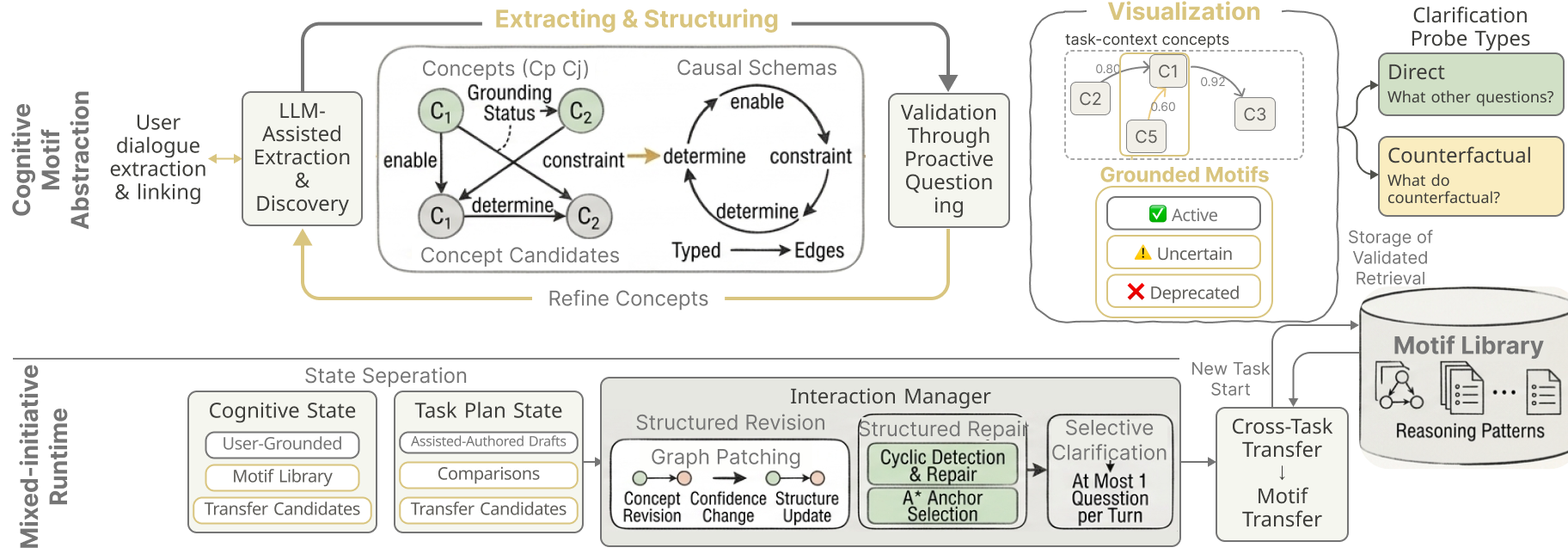}
    \caption{System framework of \Cg{}.}
    \label{fig:system}
\end{figure*}

Figure \ref{fig:system} shows the system framework of \Cg{}.

\subsection{Walkthrough Example}\label{apx:system-walkthrough}
To illustrate how \Cg{} supports bidirectional alignment throughout a task's lifecycle, we present a scenario: Emma’s 
family trip planning. 
We ground the interaction flow of \Cg{} through a scenario where Emma plans a family weekend outing, illustrating how the system facilitates cognitive alignment across different usage stages via its three interaction modes.
\emph{Stage 1: Initial Preference Gathering (Auto-revision Driven).}Emma initiates the dialogue: \textit{I want to take my family for a weekend trip nearby, ideally somewhere close to nature.''} The system automatically extracts \texttt{nature-centric} and \texttt{suburban} concepts, predicting \texttt{camping} as a high-probability candidate. During this stage, \Cg{} \textbf{auto-revises} the cognitive graph in the background, linking these preferences to a preliminary itinerary. This silent update allows Emma to continue providing unstructured input (e.g., \textit{bringing two kids and a Golden Retriever''}) while the system synchronizes constraint structures in real-time without interrupting her narrative flow.

\emph{Stage 2: Deep Logic Clarification (Clarification Driven).}When Emma mentions, \textit{My husband has been having back pain lately,''} the system detects a potential causal conflict: \texttt{back\_injury} strongly contradicts the earlier inference of \texttt{camping} (which implies sleeping bags on the ground). Since this affects the foundational plan, the runtime triggers an \textbf{uncertainty-driven clarification} probe: \textit{Given the back injury, should we stick to camping with an upgraded air mattress, or switch to a cabin hotel with nature views?''} Emma confirms, \textit{``Comfort is the priority; let's do a hotel.''} This feedback immediately instantiates a \texttt{back\_injury} $\xrightarrow{\text{determine}}$ \texttt{hotel\_accommodation} motif, resolving the inferential ambiguity.

\emph{Stage 3: Contingency-based Adjustment (User-driven Revision Driven).}
Two days before the trip, a heavy rainstorm is forecasted. Emma inputs: \textit{``It's going to rain heavily; outdoor activities are no longer viable.''} Recognizing that this upends the \texttt{nature-centric} reasoning backbone, the system does not perform an opaque overwrite. Instead, it surfaces a \textbf{user-driven revision patch}. In the graph preview, Emma sees the original outdoor items marked as \texttt{Deprecated}, replaced by system-proposed indoor alternatives like museums and aquariums. After noticing the aquarium is too far, Emma manually deletes that node via the Control Panel and clicks \textbf{Approve}. This explicit patching mechanism ensures she retains full agency over complex decision shifts.\paragraph{Stage 4: Reasoning Asset Reuse (Cross-task Transfer).}A month later, while organizing a corporate team-building event, \Cg{} retrieves patterns such as comfort-priority'' and rain-contingency'' from the \textbf{Motif Library}. It presents these as \textbf{transfer candidates}, asking: \textit{Should we apply your previous 'weather-sensitive activity' logic to this new task?''} Emma approves with one click. In this way, the cognitive logic Emma previously externalized in her personal life is transformed into a reusable intellectual template'' that assists her in professional domains.

\section{User Study}

\subsection{Demographic Information of Participants}\label{apx:participants}
Table~\ref{tab:participants} shows 12 participating users harboring diverse professions and LLM experiences. 
\begin{table*}[t]
  \centering
  \small
  \begin{threeparttable}
    \caption{Summary of demographic information of participants in the user study.}
    \label{tab:participants}
    \setlength{\tabcolsep}{5pt}
    \begin{tabular}{l c c p{3.5cm} c c p{3.2cm}}
      \toprule
      \textbf{No.} & \textbf{Gender} & \textbf{Age} & \textbf{Profession} & \textbf{Relevant Exp. (years)} & \textbf{LLM Exp. (years)} & \textbf{Frequency of LLM Use} \\
      \midrule
      P1  & Male   & 28 & Artist/designer                  & 8 & 4 & 3--5 times per week \\
      P2  & Female & 27 & Artist/designer                  & 4  & 5 & Daily \\
      P3  & Female & 25 & Programmer                       & 2  & 5 & 3--5 times per week \\
      P4  & Female & 28 & Researcher                       & 4  & 4 & Daily \\
      P5  & Female & 23 & Researcher, artist/designer      & 1  & 3 & 3--5 times per week \\
      P6  & Female & 26 & Researcher, programmer                       & 4  & 4 & 1--2 times per week \\
      P7  & Male   & 23 & Programmer                       & 5  & 3 & Daily \\
      P8  & Female & 23 & Programmer                     & 4  & 2 & 2--3 times per week \\
      P9  & Male   & 26 & Researcher                       & 6  & 4 & Daily \\
      P10 & Female & 22 & Computer science undergraduate student & 4  & 2 & 5 times per week \\
      P11 & Female & 23 & Artist/designer graduate student          & 5  & 3 & Daily \\
      P12 & Male   & 24 & Programmer          & 5  & 4 & Daily \\
      \bottomrule
    \end{tabular}
    \begin{tablenotes}
      \scriptsize
      \item Note: Relevant experience and LLM experience are reported in years.
    \end{tablenotes}
  \end{threeparttable}
\end{table*}

\subsection{Study Scenarios and Task Protocols}
\label{apx:tasks}

\subsubsection{Scenario A: Travel Planning}
This scenario examined motifs related to spatial, social, and budgetary dependencies in high-density urban environments.

\textbf{Task A1: Weekend in Hong Kong (2D/1N)}
\begin{itemize}
    \item \textbf{Initial Goal:} Design a 2-day itinerary for a first-time visitor.
    \item \textbf{Constraint Injection:} 
        (1) Sudden budget reduction; 
        (2) Emergence of "crowd-avoidance" preference; 
        (3) A travel companion joins who strictly prefers nature over city-walks.
    \item \textbf{Reasoning Focus:} Balancing urban exploration with environmental stressors and social constraints.
\end{itemize}

\textbf{Task A2: The Paris Trio (3D/2N)}
\begin{itemize}
    \item \textbf{Initial Goal:} Act as a group leader to coordinate a 3-day trip for three people with distinct personas.
    \item \textbf{Constraint Injection:} 
        (1) Realization of 3-hour wait times at major landmarks; 
        (2) Budget limits affecting accommodation choices; 
        (3) A new friend joins demanding "quiet, natural spaces" only.
    \item \textbf{Reasoning Focus:} Cross-task reuse of the "Crowd-Avoidance" and "Nature-Priority" motifs in a different geographic context.
\end{itemize}

\subsubsection{Scenario B: Personal Growth Planning}
This scenario focused on lifestyle management, examining motifs linked to temporal constraints, energy levels, and psychological momentum.

\textbf{Task B1: Adaptive Fitness Regimen}
\begin{itemize}
    \item \textbf{Initial Goal:} Create a weekly workout and nutrition plan (frequency, type, and recovery).
    \item \textbf{Constraint Injection:} 
        (1) Unstable work schedule; 
        (2) Recognition of personal procrastination patterns; 
        (3) Low energy levels during weekdays.
    \item \textbf{Reasoning Focus:} Developing motifs for "Sustainability over Intensity" to ensure habit consistency.
\end{itemize}

\textbf{Task B2: Strategic Study Planning}
\begin{itemize}
    \item \textbf{Initial Goal:} Design a learning roadmap for a new technical skill or language.
    \item \textbf{Constraint Injection:} 
        (1) Reduced daily study window; 
        (2) Motivation fluctuations (procrastination); 
        (3) Requirement for a "low-stress" learning environment to prevent burnout.
    \item \textbf{Reasoning Focus:} Transferring "Procrastination Management" and "Energy-based Scheduling" motifs from Task B1 to an academic context.
\end{itemize}



\subsection{Semi-Structure Interview Guide}\label{apx:interviewguide}
\textbf{Dimension: Externalization (Cognitive Externalization)} (RQ1 DG1): 
``Did the system help you make your reasoning visible?''
``Did the motifs help you express your thinking?''

\textbf{Dimension: Grounding (Causal Structure \& Alignment)} (RQ1 DG2): 

``Did the causal structure help the LLM understand you better than listing goals alone?'' 
``Did motif reuse feel natural?'' 
``Did the system understand your reasoning better?'' 
``Did the motif graph help you notice assumptions you hadn't explicitly stated?'' 
``How did this feel different from just listing your goals or preferences?'' 

\textbf{Dimension: Revision (Belief Revision \& Causal Propagation)} (RQ2 DG3): 
``Could you update your thinking in a principled, traceable way?'' 
``Did seeing the causal links change how you explained your reasoning to the system?'' 

\textbf{Dimension: Transfer (Cross-Task Motif Transfer)} (RQ3 · DG4): 
``Did reusing reasoning structures across tasks feel natural and useful?''

\textbf{Dimension: Trust / Control (Perceived Alignment)} (RQ4): ``Did you feel the system was aligned with how you actually think?''

\subsection{A1: Self-defined Likert Scale Questionnaire} \label{apx:A1}
The following items were rated on a 7-point Likert scale (1 = Strongly Disagree, 7 = Strongly Agree).

\textbf{Dimension: Diagnosis (Cognitive Failure Diagnosability)}
\begin{itemize}
    \item[\textbf{Q1.}] When the AI's output does not meet my expectations, I can clearly identify which specific thought or assumption went wrong.
    \item[\textbf{Q2.}] I can distinguish between cases where "the AI misunderstood my intent" and where "my own thoughts were originally unclear."
    \item[\textbf{Q3.}] When I discover an issue with the output, I know exactly where to begin making revisions.
\end{itemize}

\textbf{Dimension: Externalization (Cognitive Externalization)}
\begin{itemize}
    \item[\textbf{Q4.}] The system helped me express the reasoning behind my decisions, rather than just the final desired outcome.
    \item[\textbf{Q5.}] I was able to see the dependencies among different ideas (e.g., assumptions and constraints) and how they are interrelated.
    \item[\textbf{Q6.}] Using this system made me more aware of my own thinking process during the task.
\end{itemize}

\textbf{Dimension: Grounding (Causal Structure \& Alignment)}
\begin{itemize}
    \item[\textbf{Q7.}] The system understood the reasons behind my preferences, not just the preferences themselves.
    \item[\textbf{Q8.}] I felt that the LLM's responses were grounded in my actual reasoning, not just surface-level requests.
\end{itemize}

\textbf{Dimension: Revision (Belief Revision \& Causal Propagation)}
\begin{itemize}
    \item[\textbf{Q9.}] When I changed my mind, the system helped me update my reasoning in a structured way.
    \item[\textbf{Q10.}] I could track how my decisions evolved throughout the conversation.
    \item[\textbf{Q11.}] When I revise a specific idea, I naturally consider which other related decisions it will affect.
    \item[\textbf{Q12.}] After modifying a single condition, I know which previously made judgments need to be reconsidered.
\end{itemize}

\textbf{Dimension: Transfer (Cross-Task Motif Transfer)}
\begin{itemize}
    \item[\textbf{Q13.}] The reasoning patterns from the first task helped me get started more quickly on the second task.
    \item[\textbf{Q14.}] The way the system carried prior reasoning into the second task felt relevant to the new task context.
\end{itemize}

\textbf{Dimension: Trust/Control (Perceived Alignment)}
\begin{itemize}
    \item[\textbf{Q15.}] I trusted that the system accurately represented my reasoning.
    \item[\textbf{Q16.}] I felt in control of how my reasoning was captured and used by the system.
    \item[\textbf{Q17.}] Interacting with this system felt different from simply stating my goals to an LLM.
\end{itemize}

\section{Full System Log Visualizations}
\label{sec:appendix-systemlogs}

\begin{figure*}[t]
    \centering
    \includegraphics[width=\textwidth]{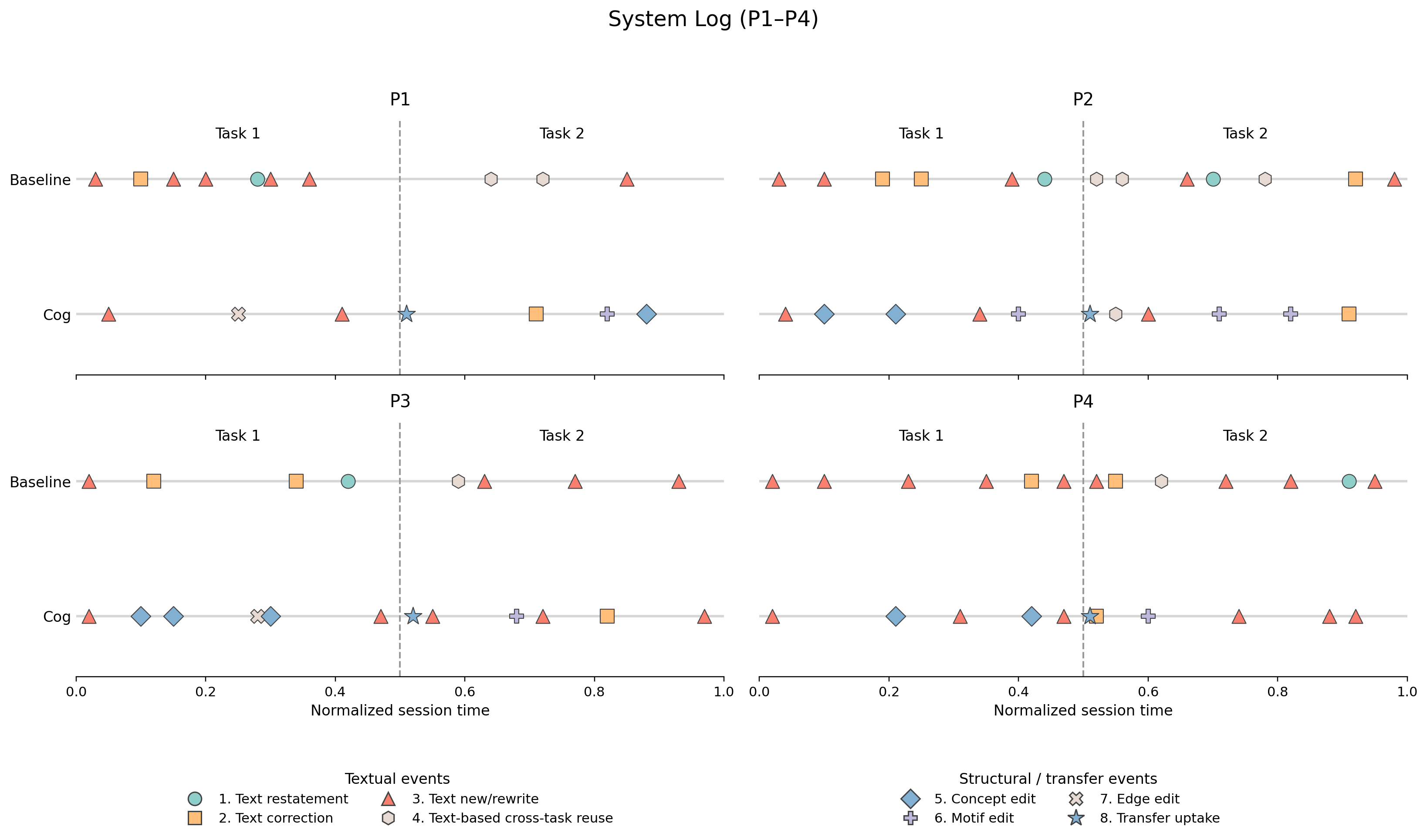}
    \caption{Full system-log timelines for Participants P1--P4. These traces mainly illustrate lighter structural uptake and several text-dominant trajectories, complementing the findings that some participants relied on \Cg{} primarily as an inspectable coordination surface rather than as a heavily edited graph. Each panel shows normalized interaction timelines under Baseline and \Cg{}, with a Task~1/Task~2 boundary. Events 1--4 correspond to text-based maintenance and reuse; Events 5--8 represent structural edits and transfer uptake in \Cg{}.}
    \Description{A grid of four panels showing system-log timelines for participants P1 to P4. Each panel includes two horizontal rows labeled Baseline and CogInstrument, separated by a vertical dashed line marking the Task 1 and Task 2 boundary. Colored markers indicate eight event types. Baseline rows contain only text-based events, while CogInstrument rows show occasional structural edits and limited transfer uptake, illustrating relatively light structural adoption and several text-dominant interaction styles.}
    \label{fig:appendix-tb1}
\end{figure*}

\begin{figure*}[t]
    \centering
    \includegraphics[width=\textwidth]{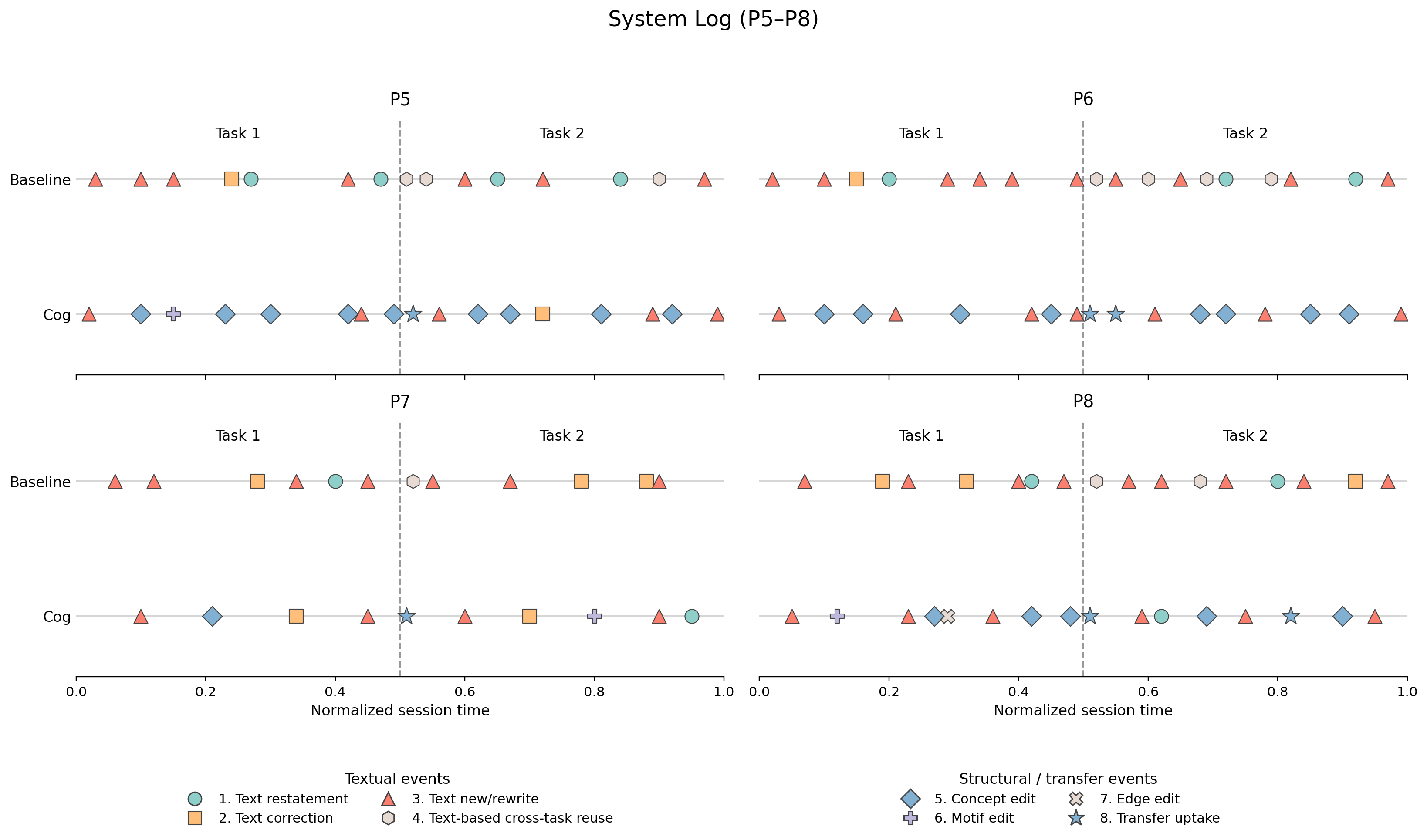}
    \caption{Full system-log timelines for Participants P5--P8. Compared with Figure~\ref{fig:appendix-tb1}, these participants show stronger adoption of structure-level interaction in \Cg{}, including repeated concept edits, motif edits, and selective transfer uptake. The traces support the finding that many users shifted from early text grounding to more localized structural repair once an initial scaffold had formed.}
    \Description{A grid of four panels showing system-log timelines for participants P5 to P8. Each panel includes Baseline and CogInstrument rows with a dashed boundary between Task 1 and Task 2. Colored markers indicate eight event types. Compared with the previous figure, the CogInstrument rows show more frequent concept edits, motif edits, and occasional transfer uptake, reflecting stronger adoption of structure-level revision after initial text-based grounding.}
    \label{fig:appendix-tb2}
\end{figure*}

\begin{figure*}[t]
    \centering
    \includegraphics[width=\textwidth]{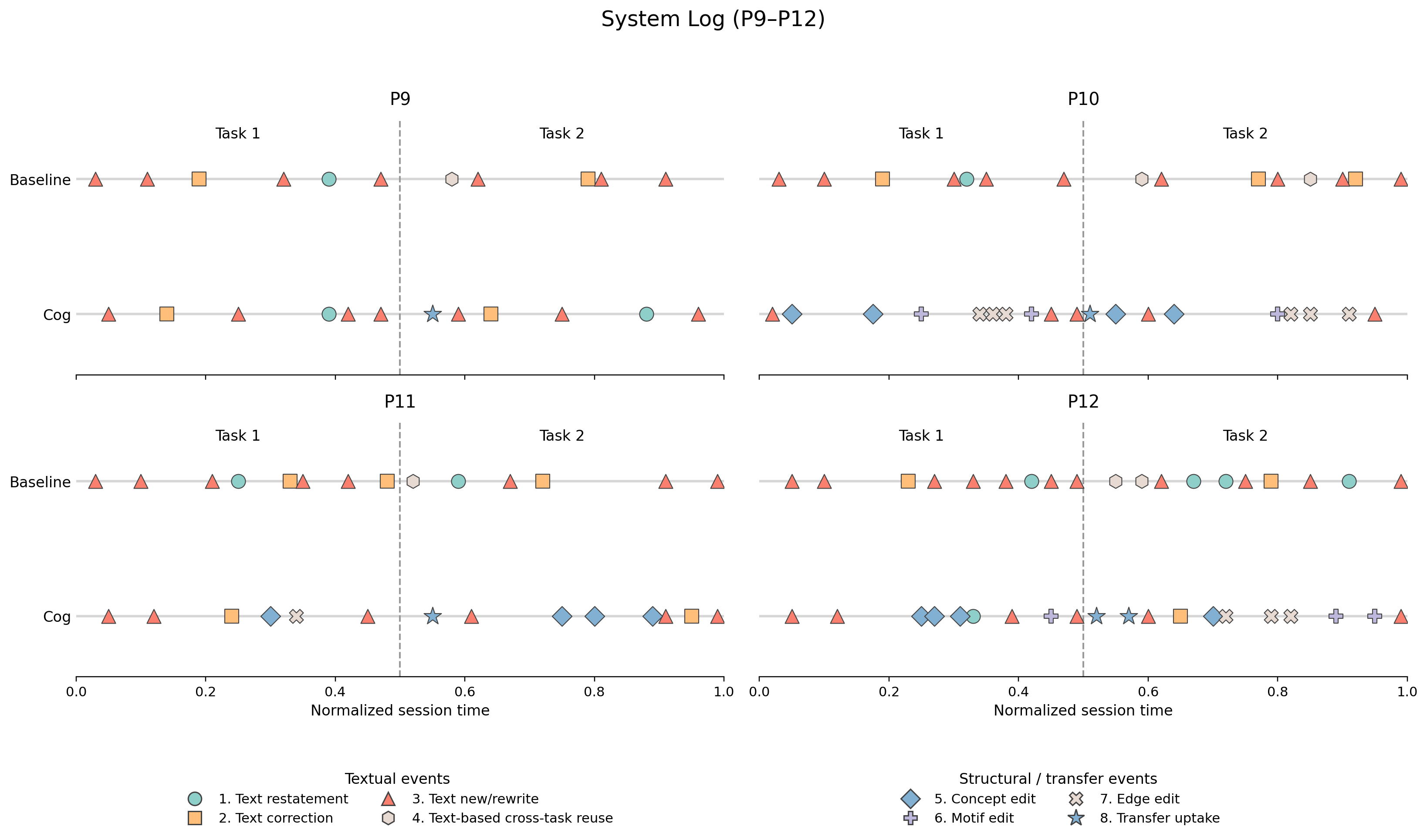}
    \caption{Full system-log timelines for Participants P9--P12. These traces highlight the broadest range of strategies in the corpus, including text-dominant interaction, mixed-mode revision, and exploratory use of structural and transfer mechanisms. Together with Figures~\ref{fig:appendix-tb1} and \ref{fig:appendix-tb2}, they show that the benefits of \Cg{} emerged through heterogeneous patterns of inspection, patch revision, and selective reuse rather than through uniform heavy graph editing.}
    \Description{A grid of four panels showing system-log timelines for participants P9 to P12. Each panel contains Baseline and CogInstrument rows separated by a Task 1 and Task 2 boundary. Some participants remain largely text-dominant, while others combine text interaction with concept edits, motif edits, edge edits, and transfer uptake. The figure illustrates substantial variation in how participants engaged with the system’s structural and transfer affordances.}
    \label{fig:appendix-tb3}
\end{figure*}

To provide full transparency into the interaction timelines underlying our log analysis, we include complete system-log visualizations for all 12 participants in Figures~\ref{fig:appendix-tb1}--\ref{fig:appendix-tb3}. These figures extend the representative examples shown in Figure~\ref{fig:likert-constructs} by exposing the full participant-level distribution of revision styles across both conditions.

Across the full corpus, the logs show three recurring patterns that informed the findings in Section~\ref{sec:findings}. First, some participants remained largely text-dominant even in \Cg{}, using the graph primarily as an inspectable reference rather than a frequent editing surface. Second, many participants exhibited mixed-mode revision, beginning with natural-language grounding and shifting to localized concept-, motif-, or edge-level edits once a workable structure had formed. Third, a smaller subset showed stronger structural uptake, including repeated graph edits and selective transfer uptake in Task~2. These traces therefore complement the qualitative accounts by showing that the value of \Cg{} did not depend on a single usage style, but emerged through heterogeneous patterns of inspection, revision, and reuse.

Figure~\ref{fig:appendix-tb1} shows Participants P1--P4, whose traces illustrate relatively lighter structural uptake and several text-dominant trajectories. Figure~\ref{fig:appendix-tb2} shows Participants P5--P8, who more frequently adopted structure-level interaction, including concept edits and transfer uptake. Figure~\ref{fig:appendix-tb3} shows Participants P9--P12, highlighting broader variation, including text-dominant interaction, mixed-mode revision, and more exploratory use of structural and transfer mechanisms.

\section{Acknowledgment about the Use of LLM}
The authors would like to acknowledge the use of the generative AI tool in this work. Specifically, \textit{GPT-5.3} by OpenAI was utilized to: (1) assist in language refinement, including grammar and style corrections of existing manuscript text, (2) generate R code for data analysis based on our proposed analytical procedures, and (3) generate LaTeX tables from the analyzed data results. Moreover, \textit{GPT-5.3} model and \textit{text-embedding-ada-002} model API service was used through Microsoft Azure interface during system implementation. All interpretations, conclusions, and final content remain the responsibility of the authors. 
\begin{acks}
\end{acks}

\end{document}